\documentclass[11pt]{article}
\usepackage{amsmath,amsthm,latexsym,amssymb,amsfonts,epsfig}

\makeatletter
\@addtoreset{equation}{section}
\makeatother

\def\be{\begin{equation}}
\def\ee{\end{equation}}
\def\ba{\begin{eqnarray}}
\def\ea{\end{eqnarray}}
\def\Nl{{\mathchoice
{\setbox0=\hbox{$\displaystyle\rm N$}\hbox{\hbox to0pt
{\kern0.4\wd0\vrule height0.9\ht0\hss}\box0}}
{\setbox0=\hbox{$\textstyle\rm N$}\hbox{\hbox to0pt
{\kern0.4\wd0\vrule height0.9\ht0\hss}\box0}}
{\setbox0=\hbox{$\scriptstyle\rm N$}\hbox{\hbox to0pt
{\kern0.4\wd0\vrule height0.9\ht0\hss}\box0}}
{\setbox0=\hbox{$\scriptscriptstyle\rm N$}\hbox{\hbox to0pt
{\kern0.4\wd0\vrule height0.9\ht0\hss}\box0}}}}
\def\Zl{{\mathchoice
{\setbox0=\hbox{$\displaystyle\rm Z$}\hbox{\hbox to0pt
{\kern0.4\wd0\vrule height0.9\ht0\hss}\box0}}
{\setbox0=\hbox{$\textstyle\rm Z$}\hbox{\hbox to0pt
{\kern0.4\wd0\vrule height0.9\ht0\hss}\box0}}
{\setbox0=\hbox{$\scriptstyle\rm Z$}\hbox{\hbox to0pt
{\kern0.4\wd0\vrule height0.9\ht0\hss}\box0}}
{\setbox0=\hbox{$\scriptscriptstyle\rm Z$}\hbox{\hbox to0pt
{\kern0.4\wd0\vrule height0.9\ht0\hss}\box0}}}}
\def\Ql{{\mathchoice
{\setbox0=\hbox{$\displaystyle\rm Q$}\hbox{\hbox to0pt
{\kern0.4\wd0\vrule height0.9\ht0\hss}\box0}}
{\setbox0=\hbox{$\textstyle\rm Q$}\hbox{\hbox to0pt
{\kern0.4\wd0\vrule height0.9\ht0\hss}\box0}}
{\setbox0=\hbox{$\scriptstyle\rm Q$}\hbox{\hbox to0pt
{\kern0.4\wd0\vrule height0.9\ht0\hss}\box0}}
{\setbox0=\hbox{$\scriptscriptstyle\rm Q$}\hbox{\hbox to0pt
{\kern0.4\wd0\vrule height0.9\ht0\hss}\box0}}}}
\def\Rl{{\mathchoice
{\setbox0=\hbox{$\displaystyle\rm R$}\hbox{\hbox to0pt
{\kern0.4\wd0\vrule height0.9\ht0\hss}\box0}}
{\setbox0=\hbox{$\textstyle\rm R$}\hbox{\hbox to0pt
{\kern0.4\wd0\vrule height0.9\ht0\hss}\box0}}
{\setbox0=\hbox{$\scriptstyle\rm R$}\hbox{\hbox to0pt
{\kern0.4\wd0\vrule height0.9\ht0\hss}\box0}}
{\setbox0=\hbox{$\scriptscriptstyle\rm R$}\hbox{\hbox to0pt
{\kern0.4\wd0\vrule height0.9\ht0\hss}\box0}}}}
\def\Cl{{\mathchoice
{\setbox0=\hbox{$\displaystyle\rm C$}\hbox{\hbox to0pt
{\kern0.4\wd0\vrule height0.9\ht0\hss}\box0}}
{\setbox0=\hbox{$\textstyle\rm C$}\hbox{\hbox to0pt
{\kern0.4\wd0\vrule height0.9\ht0\hss}\box0}}
{\setbox0=\hbox{$\scriptstyle\rm C$}\hbox{\hbox to0pt
{\kern0.4\wd0\vrule height0.9\ht0\hss}\box0}}
{\setbox0=\hbox{$\scriptscriptstyle\rm C$}\hbox{\hbox to0pt
{\kern0.4\wd0\vrule height0.9\ht0\hss}\box0}}}}
\def\Hl{{\mathchoice
{\setbox0=\hbox{$\displaystyle\rm H$}\hbox{\hbox to0pt
{\kern0.4\wd0\vrule height0.9\ht0\hss}\box0}}
{\setbox0=\hbox{$\textstyle\rm H$}\hbox{\hbox to0pt
{\kern0.4\wd0\vrule height0.9\ht0\hss}\box0}}
{\setbox0=\hbox{$\scriptstyle\rm H$}\hbox{\hbox to0pt
{\kern0.4\wd0\vrule height0.9\ht0\hss}\box0}}
{\setbox0=\hbox{$\scriptscriptstyle\rm H$}\hbox{\hbox to0pt
{\kern0.4\wd0\vrule height0.9\ht0\hss}\box0}}}}
\def\Ol{{\mathchoice
{\setbox0=\hbox{$\displaystyle\rm O$}\hbox{\hbox to0pt
{\kern0.4\wd0\vrule height0.9\ht0\hss}\box0}}
{\setbox0=\hbox{$\textstyle\rm O$}\hbox{\hbox to0pt
{\kern0.4\wd0\vrule height0.9\ht0\hss}\box0}}
{\setbox0=\hbox{$\scriptstyle\rm O$}\hbox{\hbox to0pt
{\kern0.4\wd0\vrule height0.9\ht0\hss}\box0}}
{\setbox0=\hbox{$\scriptscriptstyle\rm O$}\hbox{\hbox to0pt
{\kern0.4\wd0\vrule height0.9\ht0\hss}\box0}}}}

\title{{\sf Lessons for Loop Quantum Gravity}\\
{\sf from Parametrised Field Theory}} 
\author{{\sf T. 
Thiemann}\thanks{{\sf 
thiemann at theorie3.physik.uni-erlangen.de,
tthiemann at perimeterinstitute.ca}}\\
\\
{\sf Inst. for Theoretical Physics III, FAU Erlangen -- N\"urnberg,}\\
{\sf Staudtstr. 7, 91058 Erlangen, Germany}\\
\\
{\sf and}\\
\\
{\sf Perimeter Institute for Theoretical Physics,}\\ 
{\sf 31 Caroline Street N, Waterloo, ON N2L 2Y5, Canada}}
\date{}

\begin{document}

\maketitle

\begin{abstract}
{\sf
In a series of seminal papers, Laddha and Varadarajan have developed in depth
the quantisation of Parametrised Field Theory (PFT) in the kind of discontinuous
representations that are employed in Loop Quantum Gravity (LQG). In one
spatial dimension (circle) PFT is very similar to the closed bosonic string
and the constraint algebra is isomorphic to two mutually commuting Witt 
algebras. Its quantisation is therefore straightforward in LQG like 
representations which by design lead to non anomalous, unitary, albeit
discontinuous representations of the spatial diffeomorphism group. 
In particular, the complete set of (distributional) solutions to the 
quantum constraints, a preferred and complete algebra of Dirac observables
and the associated physical inner product has been 
constructed.

On the other hand, the two copies of Witt algebras are classically isomorphic
to the Dirac or hypersurface deformation algebra of General Relativity
(although without structure functions). The question we address in this 
paper, also raised by Laddha and Varadarajan in their most recent paper, 
is whether we can 
quantise the Dirac algebra in such a way that its 
space of distributional solutions coincides with the one just described.
This potentially teaches us something about LQG where a classically 
equivalent formulation of the Dirac algebra in terms of spatial diffeomorphism
Lie algebras is not at our disposal. 

We find that, in order to achieve this, the Hamiltonian constraint has to 
be quantised by methods that extend those previously considered. The amount
of quantisation ambiguities is somewhat reduced but not eliminated.
We also show that the algebra of Hamiltonian constraints closes in a 
precise sense, with soft anomalies, that is, anomalies that do not cause 
inconsistencies. We elaborate on the relevance of these findings for full LQG. 
}
\end{abstract}

\newpage

\tableofcontents

\newpage

\section{Introduction}
\label{s1}

Undoubtedly the major unresloved challenge in LQG \cite{books,reviews}
is to find a proper implementation of the quantum dynamics. 
While there is a large degree of control as far as the spatial 
diffeomorphism constraint is concerned \cite{ALMMT}, 
the appropriate quantisation of the Hamiltonian constraint remains the 
hardest research problem to be solved. The ideal 
wish list comprises: 1. quantisation without anomalies, 2. faithful 
representation of the Dirac or hypersurface deformation algebra, 3.
sufficient control on the classical limit, 4. sufficient control
on the space of (distributional) solutions and the corresponding physical 
inner product and 5. lack of quantisation ambiguities. 
So far only partial fulfillment of this wishlist could 
be achieved. In \cite{QSD} an anomaly free quantisation of the Hamiltonian
constraint was proposed but the remaining issues could not be addressed.
By substituting the infinite number of Hamiltonian constraints by the single 
Master constraint
\cite{master} one cancels items 1. and 2. and makes progress on 4.
In particular, if one quantises it as a spatially diffeomorphism invariant
operator on the unique \cite{LOST} (kinematical) Hilbert space of 
LQG selected by covariance with respect to the spatial diffeomorphism
group, then one can also make make progress\footnote{In \cite{AQG}
an algebraic version of LQG was studied. However, the results obtained there
are also valid for standard LQG with minor modifications because 
a spatially diffeomorphism invariant operator must not be graph changing
which essentially leads back to the calculation performed in \cite{AQG}.
The only difference is that in \cite{AQG} subgraphs of the infinite 
graph considered may change while in LQG this is not allowed. However, these
processes are semiclassically irrelevant if one employs the coherent states
defined in \cite{Coherent,shadow}, see \cite{GHTW} for more details.}  
on 3. \cite{AQG}. 

However, despite of this, 
one may feel uneasy about the current version of the Master Constraint
which is basically the weighted squared integral of all the 
Hamiltonian constraints,
because it is possible to take anomalous Hamiltonian constraints and
still end up with a well defined operator with good semiclassical behaviour.
The anomalies express themselves in the fact the spectrum of the positive 
Master Constraint Operator has a gap. Its space of solutions is therefore 
empty unless one subtracts the gap by hand (which is finite and 
proportional to $\hbar$). While this is consistent with taking the 
semiclassical limit $\hbar\to 0$ and actually works in  several non trivial 
examples \cite{MasterTest} a better quantisation of the Hamiltonian 
constraints fulfilling items 1. and 2. is certainly desirable. Furthermore,
since many quantisations have the same semiclassical limit, fulfilling
1. and 2. could automatically reduce the amount of quantisation ambiguities 
and thus might imply progress on 5.

Several extensions of the quantisation proposed in \cite{QSD} have been 
discussed. All of these ambiguities arise because the unique 
Hilbert space representation 
is discontinuous so that the connection has to be approximated by a holonomy
along {\it some} loop. The choice of that loop and the representation that 
one takes the holonomy of label the space of ambiguities\footnote{The associated
parameter space is discrete because only the diffeomorphism equivalence 
class of the loop is important and for three valent vertices there
are no $\theta$ moduli, see \cite{books}.}, see    
e.g. in \cite{PerezAmbiguities,reviews}. For none of them, it is obvious 
that property 2. is (dis)satisfied. For in order to check it, one would 
have to compute the commutator between two Hamiltonian constraints 
on the kinematical Hilber space and to decide whether the resulting object
is a quantisation of the right hand side of the corresponding classical 
Poisson bracket. This turns out to be very difficult for three independent 
reasons: A. The classical right hand side involves an infinitesimal spatial
diffeomorphism constraint, whose quantum analog does not exist. B. In order 
to avoid the anomaly, those operators are chosen as graph changing\footnote{
That is, the 
loop in question is never contained in the graph considered.} but to date 
no semiclassical states have been constructed with respect to which graph
changing operators have a good semiclassical limit. C. If one recalls the 
algebraic manipulations that one has to perform in the classical calculation
of the Poisson brackets, then it is clear that one can repeat them quantum 
mechanically only semiclassically, however, no semiclassical states are
available as already mentioned.

It is therefore very difficult to decide whether any of the operators proposed
leads us into the right direction. It is at this point where input from 
non trivial toy models, that allow for a complete solution, may guide us 
further and one such model is parametrised field theory (PFT) in two
spacetime dimensions. In general, PFT is a free field theory 
involving one or more scalar fields $\phi$ on a flat, fixed background 
spacetime 
$(M,\eta)$ of Minkowski signature and of any dimension which one makes 
diffeomorphism invariant by pulling back the sclar field and the background 
metric by an arbitrary spacetime diffeomorphism $X$. The resulting action
now also depends on $X$ as well and reduces in the gauge $X={\rm id}$ to the 
original action in adapted (Cartesian) coordinates. PFT is therefore a 
diffeomorphism
invariant theory much like GR\footnote{One may be puzzled by the fact that
PFT is at the same time diffeomorphism invariant and backgrond dependent.
This happens because the metric is here not considered as a dynamical field.}
and thus serves as an interesting testing ground for the many technical 
and conceptual issues of full fledged Quantum Gravity as has been stressed 
and worked out in a series of seminal papers by Kucha{\v r}
\cite{Kuchar}.
In particular, since 
we know a bona fide quantisation of the model in the gauge $X={\rm id}$ 
(Fock representation) it appears to be a trivially solvable theory.
Surprisingly, things are not that trivial as pointed out by Torre and 
Varadarajan in \cite{Torre}: The Fock representations for flat and 
curved embdeddings of spatial slices in $M$ for $\dim(M)>2$ are in general 
unitarily
inequivalent. In other words, different, classically perfectly equivalent 
gauges lead to unitarily inequivalent QFT's! Beautifully, by treating
the embedding variables as dynamical fields and applying to them
LQG like representations while Fock representations for the scalar
fields are kept, this obstruction 
can be overcome \cite{Varadarajan0}. Moreover, the quantum constraint reduction 
leads to a theory unitarily equivalent to the usual Fock representation
in the gauge $X={\rm id}$.

A natural question is therefore to ask, what kind of QFT would result if
one did not fix a gauge but would rather treat the system \'a la Dirac
as a diffeormorphism invariant theory which in the canonical framework thus 
leads to spatial diffeomorphism and Hamiltonian constraints. Obviously,
given the full arsenal of techniques that have been developed for LQG,
it is natural to apply LQG methods to quantise the whole system 
(i.e embedding variables and scalar field) which in turn is 
the reason for why it is an interesting model for LQG because we know 
in principle the full solution. As expected from the purely algebraic
(or geometric, i.e. action independent) proof in \cite{Hojman}, the 
Poisson algebra of the constraints is the Dirac or hypersurface deformation 
algebra. Despite the fact that the gauge fixed theory is free, the unfixed
theory is interacting and the Dirac algebra closes with non trivial 
structure functions only {\it unless we are in two spacetime dimensions}. 
The case $D=2$ therefore leads to a further simplification, namely 
the hypersurface deformation algebra is a true (albeit infinite dimensional) 
Lie algebra, a fact that is being exploited by a close relative of 
2D PFT, namley the bosonic string \cite{GSW}.

In particular, if $M\cong \mathbb{R}\times S^1$ just 
as in closed string theory, it is possible to switch from the Dirac algebra 
to a classically equivalent Lie algebra which is simply the direct sum of 
two spatial diffeomorphism algebras for $S^1$. This fact and the fact that 
the LQG
representation by design is well adapted to spatial diffeomorphims asked 
for a quatisation of the closed bosonic string by LQG methods 
\cite{LQGString}. Similarly, in 2D PFT one may exploit this fact and 
completely solve the theory. This has been done in great detail in  
impressive works by Laddha and Varadarajan \cite{Varadarajan}.  
However, in full LQG in 4D this ``trick''
is not at our disposal and thus in order to serve as a true testing 
ground for LQG one should not solve 2D PFT using it but using the original
Dirac algebra. Yet, in contrast to 4D LQG, we know in 2D PFT what the 
answer must be and thus 2D PFT may serve as a guideline for how to 
faithfully represent the 4D Dirac algebra in the LQG representation. \\
\\
This presents a real challenge:\\
{\bf If we do not manage to quantise the Dirac algebra 
for 2D PFT as to yield the known and correct result as given to us by the 
miracle that happens in 2D, then how can we hope for the correct quantisation
of the Dirac algebra of 4D LQG which is much more complicated and involves 
non trivial structure functions?}\\
\\
This is the basic question that we analyse in this paper:\\
Is it possible to find a quantisation of the spatial diffeomorphism and 
Hamiltonian constraints respectively for closed 2D PFT using LQG techniques 
such that they (rather their algebraic
dual) annihilate the solutions to the two, classically equivalent, spatial
diffeomorphism constraints? This important question has been formulated for the 
first time in the 
papers \cite{Varadarajan} where partial answers were announced.\\
Fortunately, the answer is affirmative. Surprisingly, however, as also 
has been announced in \cite{Varadarajan}, the 
quantisation of the Dirac algebra is non trivial in the sense that it uses 
techniques so far not considered in \cite{QSD} and their relatives. This    
strengthens the suspicion that the techniques of \cite{QSD} 
should be generalised.\\
\\
The architecture of this paper is as follows:\\
\\
In section 2 we review closed 2D PFT following closely \cite{Varadarajan}. 
Our treatment will be much less complete than \cite{Varadarajan} 
and we will simplify the discussion where possible.
We urge the careful reader to refer to \cite{Varadarajan} for all the 
missing details. Notice, however, that
we consider a quantisation slightly different from the one employed in 
\cite{Varadarajan} which is technically somewhat simpler and does not 
qualitatively affect the main topic of the present article.

In section 3, which contains the main result of our work, we find 
suitable quantisations of the spatial diffeomorphism and Hamiltonian
constraints which annihilate the space of solutions to the classically
equivalent two copies of spatial diffeomorphism constraints. 
Here we follow to some extent the same route as in 4D LQG \cite{QSD},
in particular we consider density one valued operator valued distributions,
as these are the only ones that have a chance to be quantised in LQG
like representations as was shown in \cite{QSD}. Their classical expression
(using as in LQG the volume) was already sketched in \cite{Varadarajan}.
Their 
constraint algebra closes by inspection and we show that there is 
a precise correspondence between the classical hypersurface algebra
and the quantum version, including a soft anomaly which however does 
not render the quantisation inconsistent. 
In particular, the corresponding operators have the 
same kernel as given in \cite{Varadarajan}. As announced 
in \cite{Varadarajan}, in order to achieve this,
new regularisation techniques have to be introduced. We find in addition 
that also a non trivial renormalisation has to be performed. 

Finally, in section 4 we discuss the possible implications for 4D LQG. 
One of the most important ones is that in order to match the kernels
of the hypersurface algebra and the direct sum of the Witt algebras,
it was crucial that one {\it did} know about the reformulation in terms 
of Witt algebras, because this fact motivates to quantise a {\it different
holonomy flux like algebra} than one would consider natural from the 
point of view of the hypersurface deformation algebra. As this different
kinematical algebra and the usual one are represented discontinuously
in the quantum theory, it is not possible to represent the usual kinematical
algebra in the Hilbert space adapted to the direct sum of Witt algebras.
This observation touches on the very starting point of LQG: If similarly
in LQG one should work with a kinematical algebra that is perfectly
adapted to the quantum dynamics, then one has to completely reformulate 
LQG! There is no evidence for the emergence for such a more adapted algebra
at the moment and even if there was, the technical tools developed for LQG 
would presumably easily transferrable to the new situation. This is also 
the case for 2D PFT as we will see.

We have banned some involved calculations concerning the constraint 
algebra to an appendix.\\
\\
Finally, as communicated to the present
author, in a completely independent research carried out by Laddha and 
Varadarajan, the authors have obtained in part similar results.
Their work will be published shortly.

\section{Review of Parametrised Field Theory}
\label{s2}

In this section we collect all the formulae that we need for our limited 
purpose. See \cite{Varadarajan} for all the details. 
We separate the briefing into classical and quantum theory.
Readers who are 
familiar with \cite{Varadarajan} can safely skip this section and move 
on directly to section 3 except for our slightly different choice of 
representation in section \ref{s2.2}.

\subsection{Classical Theory}
\label{s2.1}

We consider the differentiable manifold $M=\mathbb{R}\times S^1$ together
with the flat Minkowski metric $\eta={\rm diag}(-1,1)$. 
In order to set up the 1+1 formalism we consider arbitrary foliations $X$
of $M$, i.e. one parameter families of embeddings of the circle into $M$.
Let $x^0:=t,\;
x^1:=x$ be standard time and angular variables on $\mathbb{R}$ and $S^1$
respectively. Here $t$ labels the leaves of the foliation. Then
\be \label{2.1}
X:\;M\to M;\; (t,x)\mapsto(T(t,x),X(t,x))
\ee
defines a diffeomorphism (reparametrisation). We write $X^0=T,\; X^1=X$. 
By means of $X^A,\;A=0,1$ we can pull back the flat metric to obtain
\be \label{2.2}
g=X^\ast \eta;\;\;g_{\alpha\beta}(t,x):=\eta_{AB}\;
X^A_{,\alpha}(t,x)
X^B_{,\beta}(t,x)
\ee
Given a scalar field $\phi:\;M\to \mathbb{R}$ we may also pull it back by 
$X^A$ to obtain
\be \label{2.3}
\Phi=X^\ast\phi;\;\;\Phi(t,x):=\phi(X(t,x))
\ee
Consider the free, massless scalar field action on the cylinder
\be \label{2.4}
S[\phi]=-\frac{1}{2}\int_M\;d^2X\;\eta^{AB}\phi_{,A}\phi_{,B}
\ee
and the Parametrised Field Theory (PFT) action on the cylinder 
\be \label{2.5}
S_{{\rm PFT}}[T,X,\Phi]=-\frac{1}{2}\int_M\;d^2x\;
\sqrt{|\det(g)|}\;g^{\alpha\beta}\Phi_{,\alpha}\Phi_{,\beta}
\ee
It is easy to see that (\ref{2.4}) and (\ref{2.5}) coincide. However, 
(\ref{2.5}) is reparametrisation invariant and thus is an example for a 
diffeomorphism invariant field theory although it depends on the background 
$\eta$. At the level of the Euler Lagrange equations one may check that
the field equations for $T,X$ are satisfied once those for $\phi$ are,
hence $T,X$ are gauge degrees of freedom by construction. 

Notice that by assumtion the leaves of the foliation are embedded circles 
$\Sigma_t=X(t,S^1)$ and as such $T$ is a periodic function of $x$ at fixed $t$
while $X$ is periodic modulo $2\pi R$ where $R$ is the Radius of the cylinder.
We also take $x$ to be periodic modulo $2\pi$. That is to say, $X$ and $x$
are just angle variables on the circle. 

The passage to the canonical formulation is straightforward and will not 
repeated here in much detail, see e.g. \cite{Varadarajan}. One defines 
the momenta conjugate to $T,X,\Phi$ by the functional derivatives
\be \label{2.6}
P_T(t,x):=\frac{\delta S_{{\rm PFT}}}{\delta \dot{T}(t,x)},\;\;    
P_X(t,x):=\frac{\delta S_{{\rm PFT}}}{\delta \dot{X}(t,x)},\;\;    
\Pi(t,x):=\frac{\delta S_{{\rm PFT}}}{\delta \dot{\Phi}(t,x)},\;\;    
\ee
where a dot denotes a partial derivative with respect to $t=x^0$ 
and discovers that the resulting phase space is subject to the following
constraints
\ba \label{2.7}
D &:=& P_T\;T'+P_{X}\;X'+\Pi\;\Phi'
\nonumber\\
C &:=& P_T\; X'+P_X\; T'+\frac{1}{2}(\Pi^2+[\Phi']^2)
\ea
where a prime denotes a partial derivative with respect to $x=x^1$.   
These constraints are primary, that is, the Legendre transform is singular 
and only allows to solve for $\dot{\Phi}$ but not for $\dot{T},\dot{X}$ in 
terms of the momenta.

From the explicit expressions for $P_T,\;P_X,\Pi$ one immediately sees that
they are periodic functions of $x^1$ as they depend only on $X'$.
Let us smear the constraints $D,C$ with periodic test functions 
$f:\;S^1\to \mathbb{R};\; x\mapsto f(x)$. We write for instance
\be \label{2.8}
C[f]:=\int_{S^1}\;dx\; f(x)\; C(x)
\ee
etc. and we also define the following bracket
\be \label{2.9}
[f,g]:=f'\; g-f\; g'
\ee
Then one readily computes the {\it Hypersurface Deformation Algebra 
$\mathfrak{H}$}
\ba \label{2.10}
\{D[f],D[g]\} &=& D[[f,g]]
\nonumber\\
\{D[f],C[g]\} &=& C[[f,g]]
\nonumber\\
\{C[f],C[g]\} &=& D[[f,g]]
\ea
familiar from the ADM formulation of GR. In performing those computations 
we used the periodicity of fields so that boundary terms can be dropped.
The interpretation of $D,C$ respectively is therefore that of a diffeomorphism
and Hamiltonian constraint respectively. 

One would expect the right hand side 
of the last line to depend on the inverse of the 1D metric
\be \label{2.11}
q:=g_{xx}=-[T']^2+[X']^2
\ee
However, the peculiarity of 1D is that a one form and in particular 
the derivative of a scalar field such as $T,X,\Phi$ is the same thing 
as a density of weight one and that a vector field is the same thing as 
a density of weight -1. Thus both $C,D$ are densities of weight 2 while 
the smearing test functions are densities of weight -1 in order that   
(\ref{2.8}) is meaningful. Then (\ref{2.9}) is nothing else than minus the Lie bracket between vector fields and (\ref{2.10}) makes sense as it
stands. Put differently, the 1D metric (\ref{2.11}) is a scalar density
of weight 2 and its inverse would be of weight -2. Since the integrand
of the right hand side
of the last line of (\ref{2.10}) must have overall density weight +1 
one should dedensitise $q^{-1}$ and multiply by $\sqrt{\det(q)}^2$ which gives
unity.    

The fact that $\mathfrak{H}$ is a true Lie algebra without structure functions
is a major simplification that happens only in 2D. For instance, in a 
quantisation of the constrained system \'a la Dirac one could consider 
group averaging methods in order to solve the constraints and define a 
physical inner product. Due to the structure functions, this is not possible
in higher dimensions. However, to apply group averaging techniques directly
to the system (\ref{2.7}) is not entirely straightforward. By means of the 
following canonical transformation
\be \label{2.12}
X_\pm:=T\pm X,\;P_\pm:=\frac{1}{2}(P_T\pm P_X)
\ee
and the definition 
\be \label{2.13}
Y_\pm:=\Pi\pm \Phi'
\ee
one readily computes
\be \label{2.14}
D_\pm:=\frac{1}{2}(D\pm C)=P_\pm\; X_\pm'\pm \frac{1}{4} [Y_\pm']^2
\ee
The equivalent constraints (\ref{2.14}) obey the much simpler 
{\it Diffeomorphism Algebra} $\mathfrak{D}$
\ba \label{2.15}
\{D_\pm[f],D_\pm[g]\}&=& D_\pm[[f,g]]
\nonumber\\
\{D_\pm[f],D_\mp[g]\}&=& 0
\ea
and thus generate the direct sum of two diff$(S^1)$ Lie algebras 
(Witt algebras). All of this is of course well known from string theory
and is generic to diffeomorphism invariant 2D field theories. 

The Hamiltonian flow of the Hamiltonian vector fields of $D_\pm[f_\pm]$ 
generate 
automorphisms on the phase 
space (canonical transformations) which are just the spatial diffeomorphisms
$\varphi^{f_\pm}$ generated by the vector field $f_\pm$ on $S^1$. 
Here $\varphi^{f_\pm}$ 
acts by pull back on the $\pm$ sector of the theory and leaves invariant the 
$\mp$ sector. Of course, $P_\pm,\;Y_\pm$ are densities of weight one while 
$X_\pm$ is a scalar under $\varphi_\pm$. Since $D_\pm$ mutually commute
the flow $\alpha_{\varphi^{f_+},\varphi^{f_-}}$ of $D_+[f^+]+D_-[f^-]$
results by concatenation of the actions just described in either order. 

To construct gauge invariant 
(Dirac) observables, we notice that $P_\pm$ can be eliminated via the 
constraints while $X_\pm$ are pure gauge, hence the true degrees of freedom 
can be identified with $Y_\pm$. Therefore we can proceed as in \cite{GHTW}
and consider the gauge fixing conditions $X_\pm-\sigma_\pm=0$ corresponding
to $D_\pm$ and compute the gauge invariant extension of the scalar
$Y_\pm/X'_\pm$ off the gauge cut $X_\pm-\sigma_\pm$. The result is 
\be \label{2.16}
O_{Y_\pm}(\sigma_\pm)=[\frac{Y_\pm}{X'_\pm}(x)]_{X_{\pm}(x)=\sigma_\pm}=
\int\; dx\; Y_\pm\;\delta(X_\pm-\sigma_\pm)
\ee
where the $\delta$ distribution is periodic modulo $2\pi R$. Alternatively
we can integrate (\ref{2.16}) against the Fourier modes $\exp(in\sigma_\pm/R)$
to arrive at the Fourier coefficients
\be \label{2.17}
O_{Y_\pm,n}=\int_{S^1}\;dx\; Y_\pm\; e^{i n X_\pm/R}
\ee
with $n\in \mathbb{Z}$ also considered in \cite{Varadarajan}.

\subsection{Quantum Theory}
\label{s2.2}

The point of recalling all of these well known facts is that (\ref{2.10}) 
or (\ref{2.15})
bring us into a situation very close to LQG in 4D. We have a constrained 
Hamiltonian system part of whose constraint algebra generates spatial 
diffeomorphisms. Therefore one naturally can apply LQG quantisation 
techniques and one would first of all consider a kinematical Hilbert
space representation of the Weyl algebra determined by the phase space
with respect to which the spatial diffeomorphism group is implemented
unitarily and without anomalies similar to \cite{LOST}. Then one can 
apply group averaging techniques in order to solve the spatial 
diffeomrophism constraints and construct a Hilbert space of spatially 
diffeomorphism invariant states. With respect to the system (\ref{2.10}) one 
would then still be left with the scalar constraints and one could try
to define it as in \cite{QSD}. However, given the reformulation 
(\ref{2.15}) it is much more convenient to consider $P_\pm,X_\pm,Y_\pm$
as the elementary variables\footnote{From $Y_\pm$ we can reconstruct 
$\Phi$ only up to a constant. This zero mode however decouples from the 
constraints \cite{Varadarajan} and will therefore not be considered in this paper.} 
and to use 
the constraints (\ref{2.15}) because, in a sense, we now have two commuting
spatial diffeomorphism groups and we can apply the LQG methods to both 
of them separately. Then, after solving both diffeomorphism constraints, no 
scalar constraint is left and one arrives at a {\it complete solution of the 
theory!} This, and much more has been done in the seminal work 
\cite{Varadarajan}.

As already mentioned in the outlook part of \cite{Varadarajan}, given
this complete solution, it would now be very interesting to go back 
to the original system (\ref{2.10}), to quantise it by following
the steps of \cite{ALMMT,QSD} and to compare with the results already
obtained. In particular, one would like to see whether there is
a quantisation of (\ref{2.10}) such that the corresponding 
dual operators annihilate the kernel of (\ref{2.15}). This is what
we will do in the next section. In the present section we just recall
the elements from \cite{Varadarajan} that we need. We will, however, deviate
somewhat in the precise technical implementation from \cite{Varadarajan}
as we will indicate explicitly. \\
\\
The classical phase space consists of the embedding sector described by
the variables $(X_\pm,P_\pm)$ and the matter sector described by the 
variables $Y_\pm$. The embedding sector is gravity like, hence we use an
LQG like representation \cite{LOST} 
for which the $X_\pm(x)$ (``Ein-Bein'') and
the $\exp(ik P_\pm[I])$ (``holonomy'') are well defined operators but not 
$P_\pm(x)$
itself. Here $I$ is a closed interval, $P_\pm[I]=\int_I\;dx\; P_\pm$ 
and $k\in k_0\mathbb{Z}$ where $\hbar k_0/R,\hbar k_0
\not\in 2\pi \mathbb{Q}$ is some positive
constant. The matter sector is string like, hence we choose the LQG string
representation \cite{LQGString} for which neither $Y_\pm(x)$ exist but
only the $\exp(il Y_\pm[I])$ where again $l\in l_0\mathbb{Z}$ and
$l_0,\; l_0^2\not\in 2\pi \mathbb{Q}$
is some positive constant\footnote{For simplicity
we take all quantities including $\hbar$ as dimensionless in this article.}.
These functions separate the points of the classical
phase space since the intervals $I$ can be arbitrarily ``small''. The
restrictions on $k_0,l_0$ are motivated by trying the match the
$\mathfrak{D}$ and $\mathfrak{H}$ quantisations. Notice that here we differ
somewhat from \cite{Varadarajan}: The authors there oppositely assumne
that $\hbar k_0/R=2\pi/A$ for some large positive integer and that
$l_0$ is any real number without any restriction and that
$l\in l_0\mathbb{Z}+\lambda$ where $\lambda\in \mathbb{R}$ may vary
from charge network to charge network. This leads
to certain modifications as far as the structure of the quantum observables
is concerned. We will commment on this section \ref{s4}.

Following the notation of \cite{Varadarajan} we consider graphs $\gamma$
which are arbitrary partitions of $S^1$ into disjoint open intervals $I$ 
(modulo the boundary points). Then we consider the ``charge (spin) networks''
\be \label{2.18}
T^\pm_{\gamma,k}:=\exp(i\sum_{I\in \gamma}\; k_I\; P_{\pm}[I])
\ee
and the Weyl elements
\be \label{2.19}
W^\pm_{\gamma,l}:=\exp(i\sum_{I\in \gamma}\; l_I\; Y_{\pm}[I])
\ee
The $T^\pm_{\gamma,k}$ form an Abelian algebra where the product of two
charge networks $T^\pm_{\gamma,k},\;T^\pm_{\gamma',k'}$ is the charge network
$T^\pm_{\gamma^{\prime\prime},k^{\prime\prime}}$ where 
$\gamma^{\prime\prime}$ is the coarsest partition of $S^1$ such that 
every $I\in\gamma,\;I'\in \gamma'$ is a union of intervals in 
$\gamma^{\prime\prime}$ while 
\be \label{2.20}
k^{\prime\prime}_{I^{\prime\prime}}=
\sum_{I^{\prime\prime}\subset I}\; k_I
+\sum_{I^{\prime\prime}\subset I'}\; k'_{I'}
\ee
Likewise, the 
$W^\pm_{\gamma,k}$ form a Non -- Abelian algebra where the product of two
Weyl elements is similarly defined up to a phase which follows from the 
Poisson brackets
\be \label{2.21}
\{Y_\pm[f],Y_\pm[g]\}=\pm \int_0^{2\pi}\; dx\; [f,g]=:<f,g>,\;\;
\{Y_\pm[f],Y_\mp[g]\}=0
\ee
for the smeared functions $Y_\pm[f]=\int\;dx\; f\; Y_\pm$. Care is needed 
since the characteristic functions $\chi_I$ are not smooth but rather 
$\chi_I'(x)=\delta(x,f_I)-\delta(x,b_I)$ where $b_I, f_I$ denote beginning
and final point of $I$ \cite{LQGString}. 
With the usual regularisation for the integral 
of the $\delta$ distribution over half of its support we obtain
\be \label{2.22}
<\chi_I,\chi_J>=
-[\kappa_J(f_I)-\kappa_J(b_I)
-\kappa_I(f_J)+\kappa_I(b_J)],\;\;
\kappa_I(x)=\left\{ \begin{array}{cc}
1 & x\in(b_I,f_I)\\
\frac{1}{2} & x\in \{b_I,f_I\}\\
0 & x\not\in I
\end{array}
\right.
\ee
We thus obtain 
\be \label{2.23}
W^\pm_{\gamma,l}\;W^\pm_{\gamma',l'}=
W^\pm_{\gamma^{\prime\prime},l^{\prime\prime}}
\exp(-i\frac{\hbar}{2}
\sum_{I\in \gamma,I'\in\gamma'} l_I\; l'_{I'} <\chi_I,\chi_{I'}>)
\ee
where we used the canonical quantisation rule to replace commutators 
by $i\hbar$ times the classical Poisson brackets as well as the BHC formula.
The definition of the abstract $^\ast$algebra $\mathfrak{A}$ is completed by
defining the commutation relations with the the $X_\pm(x)$
\be \label{2.24}
[X_\pm(x),T^\pm_{\gamma,k}]=\hbar[\sum_{I\in \gamma}\;k_I\; \kappa_I(x)]\;
T_{\gamma,k}
\ee
and all other commutators are zero.

There is an important subtlety, however. Since $X$ is an angular field,
$X_\pm$ is subject to the boundary condition 
$X_\pm(x+2\pi)=X_\pm(x)\pm 2\pi R$. Thus,
as a function on the circle, it is discontinuous. One way to deal with 
this is to keep explicitly track of this boundary condition in the choice 
of the 
Hilbert space representation \cite{Varadarajan}. Another possibility 
is to consider instead the continuous $S^1$ valued functions
\be \label{2.24a}
S^\pm_{\gamma,n}=\exp(i\sum_{v\in V(\gamma)} n_v X_\pm(v)/R)
\ee
with $n_v\in \mathbb{Z}$ and $V(\gamma)$ denotes the vertices of the 
graph $\gamma$. These Weyl elements still separate the points of the classical
phase space, except for the zero mode of $X$, because for instance 
\be \label{2.24b}
S^\pm_{\{v_0,v_1\},\{1,-1\}}=\exp(i[X_\pm(v_1)-X_\pm(v_0)])
=1+iX'_\pm(v_0)[v_1-v_0]+O([v_1-v_0]^2)
\ee
allows to extract $X_\pm'(x)$ as closely as we wish. 
The $S^\pm_{\gamma,n}$ are the precise analog of the point holonomies
considered for the first time in \cite{QSD} as a background independent 
algebra for 
scalar fields. Despite the fact that only integer charges are considered, 
in contrast to \cite{ALS}, almost all information about $X_\pm$
can be extracted, see also \cite{Velhinho} for similar remarks in context
of Loop Quantum Cosmology (LQC) \cite{LQC}.
The zero mode cannot be
extracted in contrast to 
\cite{Varadarajan}. Since,
however, the zero modes of both $\Phi,X$ do not play any role in 
the classical action and since anyway we consider the PFT only as a toy
model for 4D LQG that merely serves to illustrate certain technical 
constructions, we feel free to do so. Notice also that the $S_{\gamma,n}$
are the only objects needed in the construction of the observables 
(\ref{2.17}). 

In this spirit, one could consider a mathematical 
deformation of the PFT model further and 
{\it treat $X_\pm$ as periodic functions}. Doing this actually is not
PFT but it leads to certain technical simplifications which still bring us 
close to the 4D LQG situation.
In what follows, we consider both 
possibilities A. $X_\pm$ is treated as a periodic function and B. 
$X_\pm$ is not periodic but angular and we consider instead the 
$S^\pm_{\gamma,n}$. We will see that both treatments lead to qualitatively
similar results with respect to the main interest of the present 
article while concrete formulae will be slightly different.

In terms of the $S^\pm_{\gamma,n}$ the Heisenberg relations
(\ref{2.24}) are replaced by the Weyl relations
\be \label{2.24c}
S^\pm_{\gamma,n}\; T^\pm_{\gamma',k}\; S^\pm_{\gamma,-n}
=\exp(i\frac{\hbar}{R}[\sum_{v\in V(\gamma),I\in \gamma'}\;n_v\;k_I\; 
\kappa_I(v)])\; T^\pm_{\gamma',k}
\ee
The kinematical Hilbert space is simply\footnote{In \cite{Varadarajan} for 
the embedding sector a representation similar to but slightly different 
from option A is
chosen due to the different strategy to implement the discontinuity of 
$X$. For the scalar field sector the representations coincide. There one 
also imposes the zero mode constraint $Y_+([0,2\pi])-Y_-([0,2\pi])=0$ which
we ignore here.}  
\be \label{2.25}
{\cal H}={\cal H}_+\otimes{\cal H}_-,\;\;
{\cal H}_\pm={\cal H}^E_\pm\otimes {\cal H}^M_\pm
\ee
where ${\cal H}^E_\pm$ and ${\cal H}^M_\pm$ respectively are the GNS Hilbert
spaces \cite{Haag} defined by the following states on the respective algebras
\ba \label{2.26}
\omega^E_\pm(T^\pm_{\gamma',k} X_\pm(x_1)\; ..\; X_\pm(x_N)) &=&
\delta_{N,0}\; \delta_{k,0}\;\;\;\;{\rm Possibility~~A}
\nonumber\\
\omega^E_\pm(T^\pm_{\gamma',k} S^\pm_{\gamma,n}) &=&
\delta_{n,0}\; \delta_{k,0}
\;\;\;\;{\rm Possibility~~B}
\nonumber\\
\omega^M_\pm(W^\pm_{\gamma,l}) &=& \delta_{l,0}
\ea
That these are states (positive linear functionals) on the respective 
algebras follows from \cite{LOST,LQGString}. These states are 
$\mathfrak{G}:=Diff_+(S^1)\times Diff_-(S^1)$ invariant with respect to the 
automorphism 
groups on $\mathfrak{A}$ defined by the relations 
\ba \label{2.27}
\alpha_{(\varphi_+,\varphi_-)}[X_\pm(x)] &=& 
X_\pm(\varphi_\pm(x))\;\;\;\;{\rm Possibility~~A}
\nonumber\\
\alpha_{(\varphi_+,\varphi_-)}[S^\pm_{\gamma,n}] &=& 
S^\pm_{\varphi_\pm(\gamma),n}\;\;\;\;{\rm Possibility~~B}
\nonumber\\
\alpha_{(\varphi_+,\varphi_-)}[T^\pm_{\gamma,k}] &=& 
T^\pm_{\varphi_\pm(\gamma),k}
\nonumber\\
\alpha_{(\varphi_+,\varphi_-)}[W^\pm_{\gamma,l}] &=& 
W^\pm_{\varphi_\pm(\gamma),l}
\ea
whence by general theorems \cite{Haag} there is a unitary representation 
of $\mathfrak{G}$ on $\cal H$ defined by 
$U(g)\pi(a)\Omega=\pi(\alpha_g(a))\Omega$ for any $a\in \mathfrak{A}$. Here 
\be \label{2.28}
\Omega=\Omega^E_+\otimes\Omega^M_+\otimes\Omega^E_-\otimes\Omega^M_-,\;\; 
\pi=\pi^E_+\otimes\pi^M_+\otimes\pi^E_-\otimes\pi^M_-
\ee
are defined via the GNS data $({\cal H}^E_\pm,\pi^E_\pm,\Omega^E_\pm)$ 
and $({\cal H}^M_\pm,\pi^M_\pm,\Omega^M_\pm)$ induced by 
$\omega^E_\pm$ and $\omega^M_\pm$ respectively. 
One may easily check that the vector states
\be \label{2.28a}
|\gamma;k_+,l_+,K_-,l_->:=
\pi^E_+(T^+_{\gamma,k_+})\otimes
\pi^M_+(W^+_{\gamma,l_+})\otimes
\pi^E_-(T^+_{\gamma,k_-})\otimes
\pi^M_-(W^-_{\gamma,l_-})\;\;\Omega
\ee
define an ONB of ${\cal H}$ with the convention that there are no 
neighbouring intervals $I,J$ in $\gamma$ such that $k^\pm_I=k^\pm_J,\;
l^\pm_I=l^\pm+J$. Indeed, tensor products of operators on differnt graphs
applied to $\Omega$ can be written as (\ref{2.28a}) by suitably refining 
the graph and copying the charges on the refined intervals. 
One may check that the representation $U$ is not strongly continuous. 
The solution to the quantum constraints are now linear functionals $l$
defined on the dense subspace $\pi(\mathfrak{A})\Omega$ satisfying the 
constraint equations 
\be \label{2.29}
l[U(g)\pi(a)\Omega]=l[\pi(a)\Omega]
\;\;\forall\;\;a\in\mathfrak{A},\;g\in\mathfrak{G}
\ee
that is, functionals invariant under both copies of the diffeomorphism group
of $S^1$ (the loop group $S^1\mapsto S^1$). These solutions and the 
associated physical inner product as well as the action of the (exponentiated)
observables (\ref{2.16}), (\ref{2.17}) can be obtained 
explicitly using the group averaging techniques introduced in \cite{ALMMT} 
and are exibited in great detail in \cite{Varadarajan}. In particular one 
finds, due to the non compactness of $\mathfrak{G}$ (in the discrete topology)
and due to the different gauge orbit size of different elements 
$a\in \mathfrak{A}$ the same phenomenon as in LQG, namely that the physical
inner product suffers from averaging ambiguities labelled by diffeomorphism
equivalence classes $[\gamma]$ of graphs $\gamma$. The associated subspaces
are orthogonal and are superselected by the algebra of Dirac observables.

We will not need these results for what follows. For us the relation
(\ref{2.29}) will be sufficient to check whether quantisations of the 
algebra $\mathfrak{H}$ exist which annihilate the kernel of the algebra
$\mathfrak{D}$ defined by (\ref{2.29}).

\section{Quantisation of the Hypersurface Deformation Algebra}
\label{s3}

This section contains the main result of the present work. 
The strategy will be as follows:\\
From the point of view of the algebra $\mathfrak{H}$ there is no motivation 
to introduce the variables $X_\pm,P_\pm,Y_\pm$ and it would be more natural
to consider an LQG like kinematical HS based on 
1. ``fluxes'' $T(x),X(x),\Phi(x)$
and 2. ``holonomies'' of $P_T,P_X,\Pi$. However, if we did that then we would 
actually consider discontinuous representations of two different 
$\ast$algebras and therefore operators that exist in one representation do
not exist in the other already at the kinemtical level. Thus, in order 
to compare the quantisations of $\mathfrak{D}$ and $\mathfrak{H}$ we should 
keep a common kinematical representation and this should be the one 
that we reviewed in the previous section because it is well adapted to 
$\mathfrak{D}$ which in turn allows us to arrive at a complete solution.

Therefore we should also write the spatial diffeomorphism and Hamiltonian
constraint in terms of the variables $X_\pm,P_\pm,Y_\pm$ adapted to the 
chosen representation, which is achieved by inverting (\ref{2.14}), that
is
\be \label{3.1}
D=D_+ +D_-,\;\;C=D_+ -D_-
\ee
We will discuss the quantisation of $D,C$ separately.

\subsection{Spatial Diffeormorphism Constraint}
\label{s3.1}

We follow exactly the same strategy as in LQG\footnote{This was also suggested
in \cite{Varadarajan}.} and exploit the fact that 
$D$ generates a subalgebra (but not an ideal) of $\mathfrak{H}$. Computing
the Hamiltonian vector field of $D[f]$ and considering the associated 
Hamiltonian flow we obtain a canonical transformation $\beta_{\varphi^f}$
which, unsurprisingly, can be written as 
\be \label{3.2}
\beta_{\varphi_f}=\alpha_{\varphi^f,\varphi^f}
\ee
because $D[f]=D_+[f]+D_-[f]$. The associated group Diff$(S^1)$ generated
is therefore simply the {\it diagonal subgroup}
\be \label{3.3}
\{(\varphi_+,\varphi_-)\in \mathfrak{G};\;\;\varphi_+=\varphi_-\}\subset
\mathfrak{G}
\ee
The associated automorphism group $\varphi\mapsto \beta_\varphi$ then 
lifts in the same fashion to $\mathfrak{A}$ and results in a unitary
representation $\varphi\mapsto V(\varphi)$ 
of Diff$(S^1)$ on ${\cal H}$ in complete analogy as 
described in the previous section. From the point of view 
of $\mathfrak{H}$, in order that a linear functional $l$ on 
$\pi(\mathfrak{A})\Omega$ 
is in the kernel of (the dual of) $\mathfrak{H}$ it must satisfy in 
particular
\be \label{3.4}
l[V(\varphi)\pi(a)\Omega]=l[\pi(a)\Omega]\;\;\forall\;\;a\in\mathfrak{A},\;
\varphi\in {\rm Diff}(S^1)
\ee
Since $V(\varphi)=U(\varphi,\varphi)$ it is evident from (\ref{2.29}) that 
any $l$ satisfying (\ref{2.29}) also satisfies (\ref{3.4}). Hence, as expected,
there are no obstacles as far as the spatial diffeomorphism constraint is
concerned. This confirms the announcement made in \cite{Varadarajan}.

\subsection{Hamiltonian Constraint}
\label{s3.2}

Things are much more interesting with respect to the Hamiltonian 
constraint. The Hamiltonian constraints $C[f]$ do not generate a subalgebra
of $\mathfrak{H}$ and therefore do not exponentiate to a group. Therefore 
the strategy adopted in full LQG is to directly define the generator
$C[f]$ of the would be group, especially in view of the fact that in full
LQG the complete algebra $\mathfrak{H}$ is not even a Lie algebra due to
the structure functiuons involved. Even that does not work straightforwardly
because $C$ is a scalar density of weight two and as shown in \cite{QSD}
only scalar densities of weight one have a chance to be well defined 
operator valued distributions in spatially diffeomorphism covariant
representations. For the PFT considered here this is immediately obvious 
because if $C[f]=D_+[f]-D_-[f]$ could be defined as a self adjoint operator 
then we would obtain it by taking the derivative at $t=0$ of 
$U(\varphi^{tf},\varphi^{-tf})$. However, we already remarked that 
$U$ is not strongly continuous, therefore this cannot be the case. 

This suggests to follow the same route as in LQG \cite{QSD} 
and to quantise instead the function\footnote{This has been suggested also 
in \cite{Varadarajan}.}
$\tilde{C}=C/\sqrt{\det(q)}$ where 
\be \label{3.5}
\det(q)=q=g_{xx}=-[T']^2+[X']^2=-X_+' X_-'
\ee
defines the volume element of $S^1$ which is always positive for spacelike 
embeddings of the leaves of the foliation into the cylinder. In full LQG
the Hamiltonian constraint actually is naturally defined in this way 
by applying the Dirac procedure to the Einstein -- Hilbert action.
The factor $1/\sqrt{\det(q)}$ will enable us to absorb certain UV 
singularities. In 4D LQG the volume operator corresponding to the 
integral of the volume element over 3D regions plays a pivotal role in 
the definition of the Hamiltonian constraint. In analogy, we will here 
as well quantise the operator corresponding to the interval lengths
\be \label{3.6}
V(I):=\int_I\;dx\; \sqrt{|-X_+' X_-'|} 
\ee
where we have added, as in full LQG, an absolute value which is classically
allowed since classically the argument of the square root is positive anyway.

Once we have done that, there is a chance to obtain a well defined expression 
for the 
operator corresponding to 
\be \label{3.7}
\tilde{C}[f]=\int\;dx\;f\;\frac{C}{\sqrt{|-X_+' X_-'|}}
\ee
provided one manages to replace the $P_\pm,Y_\pm$ by holonomies, because 
$P_\pm,Y_\pm$ do not exist as operators. This 
is in precise analogy to the steps that are performed in full LQG \cite{QSD}
and leads to an operator that suffers from ambiguities related to the choice
of holonomies. The ambiguity is somewhat reduced on the space of solutions to 
the constraint because diffeomorphism equaivalent choices lead to the same 
kernel. Some care is needed also in the choice of ordering in order to obtain
an operator free of anomalies.

However, if one follows these exact same steps as in \cite{QSD}
it is quite obvious that the kernel defined by 
\be \label{3.8}
l[\tilde{C}[f]\pi(a)\Omega]=0\;\;\forall\;\;a\in \mathfrak{A},\;f
\ee
and (\ref{3.4}) cannot be exactly the same kernel as defined by (\ref{3.4}).
Since certainly we trust the kernel defined by (\ref{3.4}) much more
because there are no quantisation ambiguities, the question, 
first spelled out in \cite{Varadarajan}, arises  
whether the steps of \cite{QSD} can be suitably modified and refined 
in order to match with (\ref{2.29}). We will answer this question 
affirmatively in what follows. 

Before we come to the technicalities, let us sketch the procedure that we 
will follow:
\begin{itemize}
\item[i.] {\it Regularisation by Triangulation}\\
We consider a regularised operator depending on a 
triangulation of $[0,2\pi]$ and the limit of ininitely fine triangulation
corresponds to removing the regulator.
\item[ii.] {\it Inverse Volume Operator}\\
Adapting the Poisson bracket Identities developed in \cite{QSD} to the 
present situation, an inverse volume operator can be defined free of 
singularities which reduces the action of $\tilde{C}[f]$ to those 
intervals of the triangulation which contain a vertex of the graph $\gamma$
on which the vector state, on which the operator acts, depends.   
\item[iii.] {\it Quantisation of $D_\pm$}\\
The resulting expression is a sum over vertices of $\gamma$ of eigenvalues
of the inverse volume operator times a {\it double integral} of $D_\pm$
over the interval of the triangulation containg the vertex. This was precisely
the point of dividing by $\sqrt{\det(q)}$: The single integral in $C[f]$ 
which provides insufficient smearing in order to make it a well defined 
operator was replaced by a double smearing which now has a chance to result 
in a well defined operator. The key step is now to quantise these doubly
smeared $D_\pm$ in such a way, that they annihilate the solutions of
(\ref{2.29}). We do this by starting from the known action of 
$U(\varphi,\varphi^{-1})$ for a choice of diffeomorphism $\varphi$ with 
support in the given interval of the triangulation and then try to read 
off in which sense the operator $U(\varphi,\varphi^{-1})-{\rm id}$ 
can be recognised as an approximation to the doubly smeared $D_+-D_-$.
As we will see, this is indeed possible, however, the recognition 
involves {\it new quantisation elements} that have not been considered
yet in \cite{QSD}. Not only do we need new {\it regularisation} techniques but 
also a non -- trivial {\it renormalisation}.
\end{itemize}
As we see, steps i. and ii. are very similar to \cite{QSD}, however,
step iii. involves new techniques. In other words, the Poisson bracket 
identities involving the volume operator that have been exploited in 
\cite{QSD} seem to be robust, however, the simple and ambiguous choice of 
holonomy approximation to the continuum object employed in \cite{QSD} 
seems to be too naive. That the most straightforward quantisation does not
lead to the same kernel was already annonced in \cite{Varadarajan}. 
Of course, the way we obtained the ``correct'' 
holonomy approximation cannot be repeated in full 4D LQG because there 
a complete and clean solution as in 2D PFT is not available. However,
at the very least one learns that it is worthwhile considering more 
sophisticated quantisation techniques. In the conclusion we outline 
what techniques can presumably be transferred from 2D PFT to full 4D LQG
which could be directions to further research.\\
\\
We now carry out in detail the steps sketched above.

\subsubsection{Step I. Regularisation by Triangulation}

We consider the basic building blocks
\be \label{3.9}
\tilde{D}^E_\pm=\frac{P_\pm X_\pm'}{\sqrt{\det(q)}},\;
\tilde{D}^M_\pm=\frac{1}{4}\frac{Y_\pm^2}{\sqrt{\det(q)}}
\ee
whence 
\be \label{3.10}
\tilde{D}_\pm=\tilde{D}^E_\pm \pm \tilde{D}^M_\pm,\; 
\tilde{C}=\tilde{D}_+ -\tilde{D}_-,\;
\tilde{D}:=\frac{D}{\sqrt{\det(q)}}=\tilde{D}_+ +\tilde{D}_-
\ee
It therefore suffices to provide operator expressions for 
approximations of $\tilde{D}^E_\pm$
and $\tilde{D}^M_\pm$ and these will be of the same algebraic 
form for both the ``+'' and ``-'' 
sector. Thus we may drop the label $\pm$ for the purpose of this subsection.

We consider a triangulation $\tau$ of $[0,2\pi]$ into disjoint closed intervals 
$I$ (modulo boundary points). The dual partition $\tau^\ast$ is then the 
triangulation defined by the barycentres\footnote{Wrt the Euclidian metric 
on the interval $[0,2\pi]$.} $I^\ast$ of the intervals 
$I\in \tau$. If $I,J$ are neighbour intervals, $f_I=b_J$, then we set 
$I'=[I^\ast,J^\ast]$. Hence $I'\in \tau^\ast$ contains $f_I$ as an interior
point. We could, for instance, choose all intervals of the same coordinate 
length 
$\epsilon=2\pi/N$ where $N$ is the number of intervals in $\tau$.   
It follows
\ba \label{3.11}
D^E[f] &=& \lim_{\tau\to S^1} \sum_{I\in \tau} \; f(I^\ast)\;
\frac{P(I') X(\partial I)}{V(I)}
\nonumber\\
D^M[f] &=& \lim_{\tau\to S^1} \sum_{I\in \tau} \; f(I^\ast)\;
\frac{Y(I')^2}{V(I)}
\ea
where $X(\partial I)=X(f_I)-X(b_I)$. We will give the motivation to consider 
$\tau^\ast$ next to $\tau$ in a moment. 

\subsubsection{Step II. Inverse Volume Operator}

Given an interval $I$ we introduce a partition $\cal P$ of $I$ into 
intervals $J$ and want to define  
\be \label{3.12}
V(I)
=\lim_{{\cal P}\to I}\sum_{J\in {\cal P}}
\sqrt{|X_+(\partial J)\; X_-(\partial J)|}
=R\lim_{{\cal P}\to I}\sum_{J\in {\cal P}}
\sqrt{|\sin(X_+(\partial J)/R)\; \sin(X_-(\partial J)/R)|}
\ee
via the spectral theorem. The first and second expression on the right hand 
side of (\ref{3.12}) will be geared towards possibilities A and B 
respectively. Indeed, as follows from \cite{LOST}, $X_\pm(x)$ (possibility A)
or $\sin(X_\pm(x)/R)=[S^\pm_{\{x\},1}-S^\pm_{\{x\},-1}]/(2i)$ (possibility B)
is a self adjoint operator and $\pi_\pm(T^\pm_{\gamma,k_\pm})\Omega^E_\pm$
are eigenvectors with eigenvalue
\be \label{3.13}
\lambda_{\gamma,k_\pm}(x)=
\left\{ \begin{array}{cc}
\hbar\sum_{L\in \gamma}\; k^\pm_L\; \kappa_L(x) & {\rm Possibility~~A}\\
\sin(\frac{\hbar}{R}
\sum_{L\in \gamma}\; k^\pm_L\; \kappa_L(x)) & {\rm Possibility~~B}
\end{array}
\right.
\ee
The eigenvalue for $X_\pm(\partial J)$ or 
$\sin(X_\pm(\partial J)/R)$ respectively is therefore
\ba \label{3.14}
&&\lambda_{\gamma,k_\pm}(f_J)
-\lambda_{\gamma,k_\pm}(b_J)
=\hbar\sum_{L\in \gamma}\; k^\pm_L\; [\kappa_L(f_J)-\kappa_L(b_J)]\;\;
{\rm or}\;\;
\nonumber\\
&&
\sin(\lambda_{\gamma,k_\pm}(f_J)
-\lambda_{\gamma,k_\pm}(b_J))
=\sin(\frac{\hbar}{R}
\sum_{L\in \gamma}\; k^\pm_L\; [\kappa_L(f_J)-\kappa_L(b_J)])\;\;
\ea
respectively.
Since we take the limit of infinite refinement in (\ref{3.12}) we may assume
that the intervals $J\in {\cal P}$ are much smaller than the intervals 
$L\in \gamma$ of the given $\gamma$. Therefore there are three possibilities
for given $J,L$:\\
1. both $b_J,f_J$ are interior points of L\\
2. one of $b_J,f_J$ is a boundary point of $L$\\
3. either $b_J<b_L<f_J$ or $b_J<f_L<f_J$.\\
Case 1 does not give any contribution in (\ref{3.13}). Case 2 is of measure 
zero if we 
avarage over possible limits of partitions. Case 3 gives a contribution 
with average measure unity and   
$\kappa_L(f_J)-\kappa_L(b_J)=\pm 1$. It follows that the sum over $J$ in 
(\ref{3.14}) eventually reduces to those that overlap a vertex 
$v\in V(\gamma)$ of $\gamma$
and once that is the case the limit ${\cal P}\to I$ becomes trivial and we 
obtain
\ba \label{3.15}
&& V(I)|\gamma;k_+,l_+,k_-,l_->
\\
&=&
\left\{ \begin{array}{cc}
\hbar\sum_{v\in V(\gamma)\cap I}
\sqrt{|\sum_{J\in \gamma}
[k_J^+\delta_{v,b_J}-k_J^+\delta_{v,f_J}] 
[k_J^-\delta_{v,b_J}-k_J^-\delta_{v,f_J}]|}\; 
|\gamma;k_+,l_+,k_-,l_-> & {\rm (A)}\\
R\sum_{v\in V(\gamma)\cap I}
\sqrt{|\sum_{J\in \gamma}
\sin(\frac{\hbar}{R}[k_J^+\delta_{v,b_J}-k_J^+\delta_{v,f_J}])
\sin(\frac{\hbar}{R}[k_J^-\delta_{v,b_J}-k_J^-\delta_{v,f_J}])|}\; 
|\gamma;k_+,l_+,k_-,l_-> & {\rm (B)}
\end{array}
\right.
\nonumber
\ea
which in structure is very similar to full LQG just that in PFT the 
volume operator is diagonal in the charge network basis so that its 
spectrum is under full analytic control.

By inspection the volume operator has a large kernel so that its inverse is 
not densely defined. To define it we use the Poisson bracket identity
\ba \label{3.16}
&& \frac{1}{V(I)}
-\frac{X_+(\partial I)\; X_-(\partial I)}{V(I)^3}
\approx-4\frac{\{P_+(I'),V(I)\}\;\{P_-(I'),V(I)\}}{V(I)}
\\
&=& -16\{P_+(I'),V(I)^{1/2}\}\;\{P_-(I'),V(I)^{1/2}\}
=\frac{16}{k_0^2} T^+_{I',k=-k_0}\; T^-_{I',k=-k_0}
\{T^+_{I',k=k_0},V(I)^{1/2}\}\;\{T^-_{I',k=k_0},V(I)^{1/2}\}
\nonumber
\ea
which removes the Volume from the denominator and replaces the $P_\pm$ by
holonomies which in contrast to $P_\pm$ are well defind in quantum theory.
Thus, upon replacing the Poisson brackets by commutators divided by 
$i\hbar$, (\ref{3.16}) will be a densely defined operator because the 
holonomy operators are bounded. 

The non trivial step in this calculation   
is the second one: We have for arbitrary intervals $I,J$
\ba \label{3.17}
\{P_\pm(J),V(I)\} 
&=& -\frac{1}{2}\int_J\; dx\int_I\; dy\;
\frac{\{P_\pm(x),X_\pm'(y)\}\;X_\mp'(y)}{\sqrt{-X'_+(y)\; X'_-(y)}}
\nonumber\\
&=& -\frac{1}{2}\int_J\; dx\int_I\; dy\;
\frac{\delta_{,y}(x,y)\;X_\mp'(y)}{\sqrt{-X'_+(y)\; X'_-(y)}}
\nonumber\\
&=& \frac{1}{2}\int_J\; dx\int_I\; dy\;
\frac{\delta_{,x}(x,y)\;X_\mp'(y)}{\sqrt{-X'_+(y)\; X'_-(y)}}
\nonumber\\
&=& \frac{1}{2}\int_I\; dy\;
\frac{X_\mp'(y)}{\sqrt{-X'_+(y)\; X'_-(y)}}\;[\delta(f_J,y)-\delta(b_J,y)]
\nonumber\\
&=& \frac{1}{2}
[\frac{X_\mp'}{\sqrt{-X'_+\; X'_-}} \chi_I](\partial J)
\ea
Hence for $J=I'$ we have $\chi_I(f_{I'})=0,\;\chi_I(b_{I'})=1$ and thus 
\be \label{3.18}
\{P_\pm(I'),V(I)\}
=-\frac{1}{2}[\frac{X_\mp'}{\sqrt{-X'_+\; X'_-}}](b_{I'})
\approx -\frac{1}{2}\frac{X_\mp(\partial I)}{V(I)}
\ee
These approximations become exact in the limit $\tau\to S^1$ whence we may 
insert (\ref{3.16}) as an equality into the limit (\ref{3.11}).

Thus our preliminary operators corresponding to (\ref{3.11}) are written as
\ba \label{3.19}
D^E[f] &=& \lim_{\tau\to S^1} \sum_{I\in \tau} \; f(I^\ast)\;
[P(I') X(\partial I)]^\wedge\; Q(I)
\nonumber\\
D^M[f] &=& \lim_{\tau\to S^1} \sum_{I\in \tau} \; f(I^\ast)\;
[Y(I')^2]^\wedge\; Q(I)
\ea
where 
\be \label{3.20}
Q(I):=
-\frac{16}{k_0^2 \hbar^2} T^+_{I',k=-k_0}\; 
[T^+_{I',k=k_0},V(I)^{1/2}]\;
T^-_{I',k=-k_0}\; [T^-_{I',k=k_0},V(I)^{1/2}]
\ee
The quantum expressions $[.]^\wedge$ in (\ref{3.19}) remain unspecified for 
the 
moment. 

We claim in the limit $\tau\to S^1$ eventually only those $I\in \tau$
contribute to the action on $|\gamma,k_+,l_+,k_-,l_->$ 
which contain a vertex of 
$\gamma$. To see this, it is enough to remark from (\ref{3.15}) that 
$V(I)$ acts only at the vertices of $\gamma$ contained in $I$.
In fact, it acts only at the common vertices of $\gamma'_\pm$ where 
$\gamma'_\pm$ is the graph defined 
by the charges $k_\pm$. These graphs are to be distinguished from $\gamma_\pm$
which are defined by the joint\footnote{More precisely, we can 
define the graphs $\gamma^M_\pm,\;\gamma^E_\pm$ which are determined 
by $l_\pm,k_\pm$ respectively. Then $\gamma'_\pm=\gamma^E_\pm$ and 
$\gamma_\pm=\gamma^E_\pm\cup \gamma^M_\pm$ as well as $\gamma=\gamma_+\cup
\gamma_-$.} charges $k_\pm,l_\pm$. 
Now consider the outmost right factor in (\ref{3.20})
given by
\be \label{3.21}
T^-_{I',k=-k_0}[T^-_{I',k=k_0},V(I)^{1/2}] 
=V(I)^{1/2}-
T^-_{I',k=-k_0} V(I)^{1/2} T^-_{I',k=k_0} 
\ee
The first term vanishes on $|\gamma,k_+,l_+,k_-,l_->$ unless $I$ contains a 
common vertex of $\gamma'_\pm$. The second term adds new vertices $b_{I'},
f_{I'}$ to $\gamma$ before $\sqrt{V(I)}$ acts and then removes them again. Thus
$\sqrt{V(I)}$ may act non trivially at $b_{I'}$ even if $I$ does not contain
a vertex of $\gamma'_-$. However, both terms in (\ref{3.21}) 
act non trivially only if $I$ contains a vertex of $\gamma'_+$.
Since (\ref{3.20}) is a product of operators of the form (\ref{3.21})
one for the ``+'' and one for the ``-'' sector, it follows that 
the inverse volume operator only acts non trivially if in particular 
$I$ contains a common vertex of $\gamma'_+,\gamma'_-$.  

Thus indeed only those $I=I_v$ contribute containing a common vertex $v$
of $\gamma'_\pm$ which in particular is also a vertex of $\gamma_\pm$ 
because $\gamma'_\pm$ is a subgraph of $\gamma_\pm$ and   
for those $I_v$ the limit $I_v\to v$ becomes eventually trivial 
giving rise to an operator $Q(v)$ which acts only at the vertex $v$ 
and it does so diagonally. The corresponding eigenvalue 
$\lambda_{\gamma,v,k_+,k_-}$ on $|\gamma,k_+,l_+,k_-,l_->$ can be worked 
out explicitly using (\ref{3.15}), (\ref{3.20}) and (\ref{3.21}) but will 
not be needed in what follows\footnote{They are of the form 
$$
[\sqrt{\mu_v(k^v_+ +k_0,k^v_-)}-\sqrt{\mu_v(k^v_+,k^v_-)}]
[\sqrt{\mu_v(k^v_+,k^v_- +k_0)}-\sqrt{\mu_v(k^v_+,k^v_-)}]
$$
where $\mu_v(k^v_+,k^v_-)$ is the egenvalue (\ref{3.15}) at $v$ depending
on the charges $k^v_\pm$ of the two intervals $J$ with $v$ as a boundary.}. 
We can therefore summarise the discussion so
far by 
\ba \label{3.22}
D^E[f]|\gamma,k_+,l_+,k_-,l_-> &=& 
[\sum_{v\in V(\gamma)}\; f(v)\;
\lim_{\tau\to S^1} \sum_{I\in \tau,v\in I} \; 
[P(I') X(\partial I)]^\wedge\; Q(v)]\;
|\gamma,k_+,l_+,k_-,l_->
\nonumber\\
D^M[f]|\gamma,k_+,l_+,k_-,l_-> &=& 
[\sum_{v\in V(\gamma)}\; f(v)\;
\lim_{\tau\to S^1} \sum_{I\in \tau,v\in I} \; 
[Y(I')^2]^\wedge\; Q(v)]\;
|\gamma,k_+,l_+,k_-,l_->
\ea

\subsubsection{Step III. Quantisation of $D_\pm$}

So far the discussion completely parallels the construction in \cite{QSD}.
We could complete the definition of $\tilde{C}[f]$ which is assembelled 
from the building blocks (\ref{3.22}) by, for instance, 
replacing $P_\pm(I'),Y_\pm(I')$ by  
$\sin(k_0 P_\pm(I'))/k_0, \; \sin(l_0Y_\pm(I'))/l_0$ respectively which 
would result 
in a well defined operator at finite $\tau$. The limit $\tau\to S^1$ would 
exist, as in \cite{QSD} in a weak$^\ast$ operator topology that makes use
of spatially diffeomorphism invariant linear functionals (i.e. generalised
eigenstates of $V(\varphi)$ with unit eigenvalue) and would correspond 
to choosing for each graph $\gamma$ and each vertex $v$ in $\gamma$ a 
neighbourhood $I_{\gamma,v}$ containing $v$ and no other vertex of $v$. The 
choice of $I_{\gamma,v}$ is otherwise unspecified but different choices are 
equivalent in the afore mentioned topology, making the limit $\tau\to S^1$
trivial. In this sense there is much less ambiguity than in 4D LQG. 
However, there remains the representation or discretisation ambiguity, 
we could have chosen
for instance $\sin(nk_0 P_\pm(I_{\gamma,v})/(nk_0)$ for any integer $n\not=0$
and similar for $Y_\pm(I_{\gamma,v})$ (and also for the definition of $Q(v)$)
The commutator 
$[\tilde{C}[f],\tilde{C}[f']]$ between two so constructed 
Hamiltonian constraints
is not vanishing\footnote{This is because the attachment of the 
intervals $I_{\gamma,v}$ depends a priori on the full graph $\gamma$ and a 
second
action of the Hamiltonian constraint therefore depends not only on $\gamma$
but on $\gamma\cup I_{\gamma,v}$. See \cite{QSD} for more details.} 
but its dual action vanishes on spatially diffeomorphism 
invariant states because while the action of $\tilde{C}[f]$ adds new vertices
to a graph, these vertices are bounded by intervals (edges) which 
are only charged with respect to either the positive sector or the negative 
sector and these vertices are annihilated by the volume operator. 
This is because the Hamiltonian constraint does not contain additive 
terms that are products of operators from the positive and negative sector.
In other
words, the Hamiltonian constraint does not act on the vertices it creates. 
Again, this property is completely analogous to the situation in 4D LQG.

However, the dual action of the resulting $\tilde{C}[f]$ would surely
not annihilate the exact solutions of (\ref{2.29}). This may not be 
bad by itself as long as the two resulting quantum theories have the same 
classical limit. However, given the luxury of a quantisation without 
ambiguities and with the correct constraint algebra 
based on the reformulation of $\mathfrak{H}$ as 
$\mathfrak{D}$ it is of interest if there is a quantisation of $\tilde{C}$
different from this naive Wilson -- like replacement of ``connections''
$P_\pm,Y_\pm$ by holonomies $\sin(k_0 P_\pm(I)),\;\sin(l_0 Y_\pm(I))$
proposed in \cite{QSD} such that 1. (\ref{2.29}) is annihilated 
and 2. the resulting operator starts from (\ref{3.22}) and gives 
a new expression for $[.]^\wedge$. This is what we will analyse now.\\
\\
Obviously, in order that (\ref{2.29}) is annihilated, we must write 
$[.]^\wedge$ in the form $U(\varphi_+,\varphi_-)-{\rm id}$ or something
similar for some diffeomorphisms $\varphi_\pm$. 
To simplify the discussion we label states by
\be \label{3.23}
|\gamma,k'_+,l'_+,,k'_-,l'_->\equiv
|\gamma_+,k_+,l_+>\otimes |\gamma_-,k_-,l_->
\ee
with $\gamma=\gamma_+\cup \gamma_-$, where $\gamma_\pm$ are the coarsest 
graphs so that no neighbouring intervals have both the same $k_\pm$ and
the same  
$l_\pm$ charges. The $k'_\pm,l'_\pm$ then 
result by splitting edges of $\gamma_\pm$ into
those of $\gamma$. For the term corresponding
to the interval $I_{\gamma_\pm,v_\pm}$ with $v_\pm$ a vertex of 
$\gamma_\pm$ we consider diffeomorphisms
$\varphi^{\gamma_\pm,v_\pm}_\pm$ which have support in 
$I_{\gamma_\pm,v_\pm}$ (i.e. are equal to the identity outside of it)
as is motivated by the explicit expression (\ref{3.22}) and the discussion 
above. In other words, the choice of ``loop attachment'' is translated 
into a support property of the diffeomorphism. We now have 
\be \label{3.24}
U(\varphi_+^{I_{\gamma_+,v_+}}, 
\varphi_-^{I_{\gamma_-,v_-}})\;
|\gamma_+,k_+,l_+>\otimes |\gamma_-,k_-,l_->= 
|\varphi_+^{I_{\gamma_+,v_+}}(\gamma_+),k_+,l_+>\otimes 
|\varphi_-^{I_{\gamma_-,v_-}}(\gamma_-),k_-,l_->
\ee
so that we can discuss the positive and negative sector separately and 
can drop the label $\pm$ for the rest of the discussion. Notice that we can 
safely restrict the sum over vertices $v\in \gamma$ with $\gamma=\gamma_+\cup
\gamma_-$ to either $v \in \gamma_+$ or $\gamma_-$ because due to the 
operator $Q(v)$ there is non trivial action only on $v\in V(\gamma_+)\cap
V(\gamma_-)$ anyway.

Thus we should study 
\be \label{3.25}
|\varphi(\gamma),k,l>-|\gamma,k,l>
=[T_{\varphi(\gamma),k}\otimes W_{\varphi(\gamma),l}-T_{\gamma,k}\otimes
W_{\gamma,l}]\;\Omega^E\otimes \Omega^M
=[T_{\varphi(\gamma),k}T_{\gamma,-k}\otimes 
W_{\varphi(\gamma),l}W_{\gamma,-l}-{\rm id}]\;|\gamma,k,l>
\ee
for a diffeomorphism $\varphi$ with non-trivial action in some neighbourhood 
$I$ around a vertex $v$ of $\gamma$. Specifically, consider a graph 
$\gamma$ defined by $N$ vertices $v_1,..,v_N\in [0,2\pi]$ where 
$v_k<v_{k+1},\;k=1,..,N-1$ and $v_{N+1}:=v_1$ and suppose $v=v_1$ w.l.o.g.
(otherwise relabel the vertices). Then $v'_k=\varphi(v_k)=v_k$ for 
$k\not=1$ and $v'_1\not=v_1$. If we denote the edges of $\gamma$ by
$I_k=[v_k,v_{k+1}]$ then we see that $I'_k=[v'_k,v'_{k+1}]=I_k$ for 
$k=2,..,N-1$ but $I'_1=[v'_1,v_2]\not=I_1,\;I'_N=[v_N,v'_1]\not=I_N$.
If $v_1'>v_1$ then $I_1=I'_1\cup[v_1,v_1'],\; I'_N=I_N\cup [v_1,v_1']$
are disjoint decompositions,
if $v_1>v'_1$ then $I'_1=I_1\cup[v'_1,v_1],\; I_N=I'_N\cup [v'_1,v_1]$
are disjoint decompositions. We define 
$<v_1,v_1'>=[v_1,v_1']$ for $v_1<v_1'$ otherwise
$<v_1,v_1'>=[v'_1,v_1]$. 

We compute, abusing the notation\footnote{We display formulae as if 
the operators $P(I),Y(I)$ existed which is not the case. However, this 
is just for notational convenience, we could redo the same calculation 
just using the $T_{\gamma,k},\;W_{\gamma,l}$.}
\ba \label{3.26}
T_{\varphi(\gamma),k} T_{\gamma,-k} &=&
\exp(i[k_1 P(I'_1)+k_N P(I'_N)])\exp(-i[k_1 P(I_1)+k_N P(I_N)])
\nonumber\\
&=& \left\{ \begin{array}{cc}
\exp(i[-k_1 P(<v_1,v_1'>+k_N P(<v_1,v_1'>)]) & v_1<v_1'\\
\exp(i[k_1 P(<v_1,v_1'>-k_N P(<v_1,v_1'>)]) & v_1>v_1'
\end{array}
\right.
\nonumber\\
&=& \left\{ \begin{array}{cc}
\exp(-i[k_1-k_N] P(<v_1,v_1'>)) & v_1<v_1'\\
\exp(i[k_1-k_N] P(<v_1,v_1'>)) & v_1>v_1'
\end{array}
\right.
\nonumber\\
&=& 
\exp(i[k_N-k_1] P([v_1,v_1'])) 
\ea
with the understanding that $P([v_1,v_1'])=\pm P(<v_1,v_1'>)$ for 
$v_1<v_1',\;v_1>v_1'$ respectively. In what follows we just use 
orientation preserving diffeomorphisms so that $v_1'>v_1$.  

Suppose we could actually expand the exponential in (\ref{3.26}) then 
we would obtain to first order in $P$ $1+i[k_1-k_N] P([v_1,v_1'])$. 
If we compare this with the piece $P(I')X(\partial I)$ 
in (\ref{3.22}) then we see that we should choose $I'=[v_1,v_1']$
and $I=[v_1^l,v_1^r]$ with $v_1^l<v_1<v_1^r$ so that $X(\partial I)
=X(v_1^r)-X(v_1^l)$ because then 
\be \label{3.27}
X_\pm(\partial I)|\gamma,k>=\hbar(k_N-k_1)|\gamma,k>
\ee
produces precisely the same factor $k_N-k_1$ that is needed, at least
for Possibility A. Notice also
that as anticipated, $I$ has to be chosen as a neigbourhood of $v_1$ and 
$I'$ is naturally a segment in the dual of the partition containing $I$.
Hence that part of the steps leading to (\ref{3.22}) was precisely correct
and was obtained without having any input from the alternative quantization.

However, neither does $P$ in (\ref{3.22}) exist nor can we expand the 
exponential in (\ref{3.26}). Hence there is a step missing in order 
to match (\ref{3.22}) and (\ref{3.26}). A hint comes from the observation
that, for possibility A, if $P(I')$ existed and would commute with 
$X(\partial I)$ then we 
would get {\it an exact match} between (\ref{3.22}) and (\ref{3.26}) if 
we would interpret $[P(I') X(\partial I)]^\wedge$ as 
\be \label{3.28}
\exp(iP(I')X(\partial I)/\hbar)-{\rm id}
\ee
This is reminiscent to what happens in Loop Quantum Cosmology \cite{LQC}: The 
label of the ``holonomy'' is turned into an operator. Unfortunately
none of these assumtions hold. The would be operator (\ref{3.28}) has to be
written in terms of the $T_{\gamma,k}$ operators and so we have 
to somehow take $X(\partial I)$ out of the exponent. 
However, then it is now no longer
difficult to guess the correct operator expression for (\ref{3.28}).
We have for version A
\ba \label{3.29}
\exp(i[k_N-k_1] P(I'))\; |\gamma,k,l> &=& 
\sum_{k'\in k_0 \mathbb{Z}}\;
\delta_{k',k_N-k_1}\;\exp(ik'P(I')) \;|\gamma,k,l>  
\nonumber\\
&=& \sum_{k'\in k_0 \mathbb{Z}}\;\;\exp(ik'P(I'))
\delta_{k',X(\partial I)/\hbar}
 \;|\gamma,k,l>  
\nonumber\\
&=:& [\exp(iP(I')X(\partial I))]^\wedge
\ea
Here we define the Kronecker $\delta$ of an operator via the 
presentation
\be \label{3.30}
\delta_{k,k'}=\lim_{M\to \infty} \frac{1}{2M-1}\sum_{n=-(M-1)}^{M-1}
e^{in(k-k')} 
\ee
and substituting $k$ by $X(\partial I)$ yields a limit of a sum of 
operators $\exp(in(X(\partial I)-k'))$ which are well defined and unitary
since $X(x)$ is self adjoint. Notice that due to $k_0\not\in 2\pi \mathbb{Q}$ 
the sum in (\ref{3.30}) equals unity only for $k=k'$ at any finite value 
of $M$ so that the geometric sum for $k\not=k'$ is bounded by 
$[1+4/|e^{i(k-k')}-1|]/(2M-1)$. This then is the indirect but mathematically
only way to define the ``intuitively obvious'' operator 
$\exp(-iP(I')X(\partial I))$. Notice that it was to expected that an 
exponentiated form of $P(I') X(\partial I)$ had to be used, both because 
only exponentiated diffeomorphism generators and $P(I')$ exist.

For version B we proceed similarly: The only thing that needs to be replaced 
is the definition of $\delta_{k',X(\partial I)/\hbar}$ in (\ref{3.29}) 
which we now define as (since $X(\partial I)$ is not directly available)
\be \label{3.30a}
\delta_{k',X(\partial I)/\hbar}
:=\lim_{M\to \infty} \frac{1}{2M-1}\sum_{n=-(M-1)}^{M-1}
e^{-ink'\frac{\hbar}{R}}\;S_{\{v_1^l,v_1^r\},\{-n,n\}}
\ee
with the agreement that now $k_0\frac{\hbar}{R}\not \in 2\pi\mathbb{Q}$.\\
\\
We now turn to the other ingredient of (\ref{3.25}) and proceed again 
symbolically, using the same diffeomorphism as before
\ba \label{3.31}
W_{\varphi(\gamma),l}\;W_{\gamma,-l} &=& 
\exp(i\sum_{k=1}^N\; l_k Y(I'_k))\;\exp(-i\sum_{k=1}^N\; l_k Y(I_k))\;
\nonumber\\
&=&
\exp(i\sum_{k=1}^N\; l_k [Y(I'_k)-Y(I_k)])\;
\exp(\frac{1}{2}
[i\sum_{k=1}^N\; l_k Y(I'_k),-i\sum_{k=1}^N\; l_k Y(I_k)])
\nonumber\\
&=&
\exp(i[l_N-l_1] Y([v_1,v_1']))\;
\exp(\frac{1}{2}
[\sum_{k=1}^N\; l_k Y(I_k)+[l_N-l_1] Y([v_1,v_1']),\sum_{k=1}^N\; l_k Y(I_k)])
\nonumber\\
&=&
\exp(i[l_N-l_1] Y([v_1,v_1']))\;
\exp(\frac{1}{2}[l_N-l_1]\sum_{k=1}^N l_k
[Y([v_1,v_1']),Y(I_k)])
\nonumber\\
&=&
\exp(i[l_N-l_1] Y([v_1,v_1']))\;
\exp(i\sigma\frac{\hbar}{2}[l_N-l_1]\sum_{k=1}^N l_k
<\chi_{[v_1,v_1']},\chi_{I_k}>
\ea
where $\sigma=\pm$ takes care of whether we treat the positive or negative
sector. Recalling (\ref{2.22}) we see that only $I_1,I_N$ contribute 
to the sum in the exponential and we get (for $v_1<v_1'<v_2$ since $v_1'$
is displaced from $v_1$ by an arbitarily small amount)
\ba \label{3.32}
<\chi_{[v_1,v_1']},\chi_{I_1}> &=& 
-[\kappa_{I_1}(v_1')-\kappa_{I_1}(v_1)
-\kappa_{[v_1,v_1']}(f_{I_1})+\kappa_{[v_1,v_1']}(b_{I_1})]=-[1-1/2-0+1/2]=-1
\nonumber\\
<\chi_{[v_1,v_1']},\chi_{I_N}> &=& 
-[\kappa_{I_N}(v_1')-\kappa_{I_N}(v_1)
-\kappa_{[v_1,v_1']}(f_{I_N})+\kappa_{[v_1,v_1']}(b_{I_N})]=-[0-1/2-1/2+0]=1
\nonumber\\
&&
\ea
since $b_{I_1}=f_{I_N}=v_1,\;b_{I_N}=v_N<v_1,\;f_{I_1}=v_2>v_1'$. 
Thus 
\be \label{3.33}
W_{\varphi(\gamma),l}\;W_{\gamma,-l} = 
\exp(i\sigma \frac{\hbar}{2} (l_N-l_1)^2)\; \exp(i(l_N-l_1) Y([v_1,v_1']))
\ee

To relate (\ref{3.33}) to $\sigma Y(I')^2/4$ as in (\ref{3.22}) seems entirely 
hopeless at first: As we just saw, the best we can hope for is that 
(\ref{3.33})
corresponds to something like $\exp(i\sigma Y(I')^2/4)$. But neither is 
$Y(I')$ well
defined nor is there any obvious way to write it in terms of the well defined
holonomies. A hint comes from the observation that $\exp(i\sigma Y(I')^2/4)$ 
is a 
Gaussian and Gaussians are Fourier transforms of Gaussians, that is,
\be \label{3.34}
\int_{\mathbb{R}}\; dx\; e^{\pm iyx} \; e^{ik x^2}=ce^{\mp iy^2/(4k^2)},\;\;
c:=\int_{\mathbb{R}}\; dx\; e^{ikx^2}
\ee
where $c$ is a finite complex number which results from ``analytic 
continuation'' 
from imaginary values $k=ir,\;r>0$ to real values (the rigorous evaluation
uses Cauchy integral techniques). Something like formula (\ref{3.35}) 
could be applied, with the integral over $x$ replaced by a suitable sum
over $l\in l_0\mathbb{Z}$ to formally define $\exp(iY(I')^2)$ in terms 
of the $W_{\gamma,l}$ and to compare its action with (\ref{3.33}).
But even if this worked, it would not yet give the phase
$\exp(i\sigma \frac{\hbar}{2} (l_N-l_1)^2)$ in (\ref{3.33}). Yet, this 
idea turns out to be almost correct as we will see shortly.

We start by trivially rewriting (\ref{3.33}) as in (\ref{3.29})
\be \label{3.35}
W_{\varphi(\gamma),l}\;W_{\gamma,-l}\; |\gamma,k,l> 
=\sum_{l'\in l_0 \mathbb{Z}}\; \delta_{l',l_N-l_1}\;
\exp(i\sigma \frac{\hbar}{2} (l')^2)\; \exp(i l' Y([v_1,v_1']))
|\gamma,k,l>
\ee
Now we need, as in (\ref{3.29}), an operator that acts diagonally 
on $|\gamma,k,l>$ with eigenvalue $l_N-l_1$. For the embedding sector 
this was easy because this was precisly the action of $X(\partial I)$
for suitable $I$ dual to $I'=[v_1,v_1']$. But for the matter sector 
we do not have such an operator at our disposal. A hint comes from 
the following formal calculation: Suppose that $Y(I')$ was a well defined
operator. Then, using (\ref{3.32})  
\be \label{3.36}
[Y([v_1,v_1']),W_{\gamma,l}]=[Y([v_1,v_1'])-
W_{\gamma,l} Y([v_1,v_1']) W_{\gamma,l}^{-1}]W_{\gamma,l}
=-\hbar\sigma (l_N-l_1) W_{\gamma,l}
\ee
Thus, the operator ad$_{Y([v_1,v_1'])}$ would do the right thing, but of
course it does not exist. What exists is its exponential, formally
given by
\be \label{3.37}
{\rm Ad}_{W_{[v_1,v_1'],\tilde{l}}}=\exp(i\tilde{l} {\rm ad}_{Y([v_1,v_1'])})
\ee
where $W_{[v_1,v_1'],\tilde{l}}$ is the charge network with graph 
consisting of the two edges $[v_1,v_1'],\;[v_1',v_1+2\pi]$ with charges 
$\tilde{l},0$ respectively. The action of (\ref{3.37}) is given explicitly
by (similar as in Tomita -- Takesaki theory \cite{Haag})
\be \label{3.38} 
{\rm Ad}_{W_{[v_1,v_1'],\tilde{l}}} |\gamma,k,l>
=W_{[v_1,v_1'],\tilde{l}} 
T_{\gamma, k}\otimes W_{\gamma,l}
W_{[v_1,v_1'],\tilde{l}}^{-1} 
=T_{\gamma, k}\otimes W_{\gamma,l}
\exp(-i\sigma\hbar(l_N-l_1)\tilde{l})
\ee
where use was made of the fact that the GNS null space ideal is zero
for $\omega^M_\pm$ 
so that $a=[a]$ so that the algebra of charge networks {\it IS} 
(a dense subset of) the HS. 

We now combine (\ref{3.30}) and (\ref{3.38}) and write (using that 
$\hbar l_0\not\in 2\pi \mathbb{Q}$) (\ref{3.35}) as 
\ba \label{3.40}
W_{\varphi(\gamma),l}\;W_{\gamma,-l}\; |\gamma,k,l> 
&=& 
\sum_{l'\in l_0 \mathbb{Z}}\; 
\exp(i\sigma \frac{\hbar}{2} (l')^2)\; 
\exp(i l' Y([v_1,v_1']))\;
\times \nonumber\\ &&
[\lim_{M\to \infty} \frac{1}{2M-1}\sum_{n=-(M-1)}^{M-1}\;
\exp(i\sigma nl_0 l')\;
{\rm Ad}_{W_{[v_1,v_1'],nl_0}}\; 
|\gamma,k,l>
\nonumber\\
&=& 
\lim_{M\to \infty} \frac{1}{2M-1}\sum_{n=-(M-1)}^{M-1}
\sum_{l'\in l_0 \mathbb{Z}}\; 
\exp(i\sigma \frac{\hbar}{2} (l')^2)\; 
\exp(i\sigma nl_0 l')\;
\times \nonumber\\ &&
\exp(i [l'+nl_0] Y([v_1,v_1']))\;
|\gamma,k,l>\; \exp(-i nl_0 Y([v_1,v_1']))\;
\nonumber\\
&=& 
[\sum_{l'\in l_0 \mathbb{Z}}\; 
\exp(i\sigma \frac{\hbar}{2} (l')^2)\; 
\exp(i l' Y([v_1,v_1']))]\;
|\gamma,k,l>\; 
\times \nonumber\\ &&
[\lim_{M\to \infty} \frac{1}{2M-1}\sum_{n=-(M-1)}^{M-1}
\exp(-i\sigma \frac{\hbar}{2} (nl_0)^2)\; 
\exp(-i nl_0 Y([v_1,v_1']))]
\ea
where the phase $\exp(i\sigma nl_0 l')$ in the first line of (\ref{3.40})
had to be included so that the action of 
${\rm Ad}_{W_{[v_1,v_1'],nl_0}}$ combines to an effective 
$\exp(inl_0\sigma(l'-(l_N-l_1))$ as desired and this phase could be absorbed
into the $l'$ summation by correcting for a phase 
$\exp(-i\sigma \frac{\hbar}{2} (nl_0)^2)$ in the last step. 
In the second step we formally interchanged the sums which is allowed
at finite $M$ so that keeping finite $M$ and take the limit only
at the end has to be considered as a regularisation.

Let us define the operator 
\be \label{3.41}
[\exp(-i\sigma Y(I')^2/4)]_M^\wedge:=\sum_{l\in l_0\mathbb{Z};\;|l|\le 
(M-1)l_0} \; e^{i\sigma\hbar l^2} \; W_{I',l} 
\ee
with $I'=[v_1,v_1']$. Then (\ref{3.42}) may be written as 
\be \label{3.42}
W_{\varphi(\gamma),l}\;W_{\gamma,-l}\; |\gamma,k,l> 
\lim_{M_1,M_2\to \infty} \frac{1}{2 (M_2-1)}
\;[\exp(-i\sigma Y(I')^2/4)]_{M_1}^\wedge \; |\gamma,k,l>
[[\exp(-i\sigma Y(I')^2/4)]_{M_2}^\wedge]^\ast
\ee
Comparing with our initial idea (\ref{3.34}) we see that our guess was 
close to the final result, however, there are three non trivial differences:
\begin{itemize}
\item[1.] The integral in (\ref{3.34}) was replaced by an infinite sum. 
Using the Poisson resummation formula, the naive Gaussian is thus replaced by 
its periodification (with period $2\pi/l_0$)
\be \label{3.43}
\sum_l e^{i\sigma l^2} e^{ily}\propto \sum_l e^{-i\sigma (y+l)^2/4}
\ee
which is quite unexpected.
\item[2.] One of the limits $M_1,M_2\to\infty$ is accompanied by an 
infinite renormalisation factor $\propto 1/M_2$. 
\item[3.] Up to this renormalisation factor, the action of 
$[\exp(-i\sigma Y(I')^2/4)]_{\infty}^\wedge$ is by ``conjugation''
of $|\gamma,k,l>=T_{\gamma,k}\otimes W_{\gamma,l}$ rather than 
simple action from the left. This is different from the embedding sector and 
might seem unusal at first sight. However, it is actually not as we will now 
explain.
\end{itemize}
The action of the classical diffeomorphism $\varphi^v\in{\rm Diff}_\pm(S^1)$ 
defined by the 
vector field $v$ on the phase space is determined by its Hamiltonian flow
on the Poisson algebra $\mathfrak{P}$ of functions  
\be \label{3.44}
\alpha^\pm_{\varphi^v}(a)=\exp(\chi_{D_\pm[v]})\cdot a=\sum_{n=0}^\infty\; 
\frac{1}{n!} \{D_\pm[v],a\}_{(n)}
\ee
with the iterated Poisson bracket $\{b,a\}_{(n+1)}=\{a,\{b,a\}_{(n)}\},\;
\{b,a\}_{(0)}=a$. Upon canonical quantisation, one replaces the classical 
Poisson algebra by the associated $^\ast-$algebra $\mathfrak{A}$ which 
arises from the canonical quantisation rule, that is, from 
replacing Poisson brackets by commutators divided by $i\hbar$. Applied 
to (\ref{3.44}) this means (denoting elements of $\mathfrak{P}$ and 
$\mathfrak{A}$ by the same letters)
\be \label{3.44a}
\alpha^\pm_{\varphi^v}(a)=\sum_{n=0}^\infty\; 
\frac{(i\hbar)^{-n}}{n!} [D_\pm[v],a]_{(n)}=
\exp(\frac{D_\pm[v]}{i\hbar})\;a\exp(-\frac{D_\pm[v]}{i\hbar})
\ee
provided $D[v]$ exists as an operator in a given representation. We 
see that the quantum automorphism $\alpha^\pm_{\varphi^v}$ naturally
acts by conjugation on $\mathfrak{A}$ 
by the would be unitary operators 
$\tilde{U}_\pm(\varphi^v)=\exp(-iD_\pm[v]/\hbar)$.
However, even if $D_\pm[v]$ and thus $\tilde{U}_\pm$ 
does not exist, it is still possible to define 
$\alpha^\pm_{\varphi^v}$ on $\mathfrak{A}$ simply by lifting the geometric 
action of $\varphi^v$ via pull-back from $\mathfrak{P}$ to $\mathfrak{A}$.

Now the representation we are considering is the GNS representation 
defined by a Diff$_\pm(S^1)$ invariant 
state $\omega_\pm=\omega^E_\pm\otimes \omega^M_\pm$ 
on $\mathfrak{A}$ and after dividing out the GNS null ideal it is enough to 
consider the subalgebra generated by 
$T^\pm_{\gamma,k}\otimes W^\pm_{\gamma,l}$ which {\it defines} a dense subspace 
in the corresponding Hilbert space ${\cal H}_\pm$. We can therefore drop 
the brackets $[.]$ defining the corresponding GNS equivalence class and 
identify elements $a$ of that subalgebra with vector states in the 
corresponding GNS Hilbert space. As is well known \cite{Haag}, in this 
representation 
the {\it outer} automorphisms $\alpha^\pm_{\varphi^v}$ define rigorously 
unitary operators
$W_\pm(\varphi^v)$ via 
$\alpha^\pm_{\varphi^v}(a)=:W_\pm(\varphi^v)\cdot a$. If these outer 
automorphisms are also {\it inner}, then we can find unitary operators
$U_\pm(\varphi)$ constructed (possibly as limits) from $\mathfrak{A}$ and so 
we see that the heuristic $\tilde{U}_\pm$ should be identified with the 
rigorous $U_\pm$ whenever both exist. 

The point to notice, however, is that if $U_\pm$ exists then it 
acts on the ${\cal H}$ 
(the completion of the linear span by the above subalgebra of 
$\mathfrak{A}$ in the inner product defined by 
$<b,a>_\pm=\omega_\pm(b^\ast a)$) by conjugation and not by action from the 
left.
It is the unitary operator 
\be \label{3.45}
W_\pm(\varphi)\cdot a:=\alpha^\pm_\varphi(a)
=U_\pm(\varphi)\; a\; U_\pm(\varphi)^{-1}
\ee
that we are representing on ${\cal H}$ by action from the left and not 
$U_\pm(\varphi)$. This is always the case in GNS representations. 
If the action of $W_\pm$ were continuous, it would define a generator 
$D'_\pm[v]$ which however could in general not have much to do with 
$D_\pm[v]$ as is well known from Tomita -- Takesaki theory \cite{Haag}.
In fact, it would be $D'_\pm[v]\cdot\propto [D_\pm[v],.]$ if the latter
existed. 
If, however, $U_\pm(\varphi)\cdot 1=1$ i.e. if the GNS vacuum vector 
$\Omega=1$ 
is invariant under $U_\pm$ (it is obviously so under $W_\pm$) then 
$W_\pm=U_\pm$ because 
\be \label{3.46}
W_\pm(\varphi)\cdot a=
U_\pm(\varphi)\cdot a\cdot U_\pm(\varphi)^{-1}
=U_\pm(\varphi)\cdot a\cdot U_\pm(\varphi)^{-1} \cdot 1
=U_\pm(\varphi)\cdot a
\ee
There is no contradiction with our experience with ordinary quantum 
mechanics, for there 
one is usually given the following situation: \\
By the Stone -- von Neumann theorem, the only irreducible and continuous 
(with respect to the Weyl algebra $\mathfrak{A}$) representation is the 
usual Schr\"odinger
representation $\pi$ on ${\cal H}=L_2(\mathbb{R},dx)$. Suppose
a given Hamiltonian $H$ has a cyclic and separating\footnote{The equation
$\pi(a)\Omega=0$ has only the trivial solution $a=0$.} (for the Weyl algebra)
ground vector state $\Omega,\;||\Omega||=1$, that is, $H\Omega=0$. 
We have then inner automorphisms $\pi(\alpha_t(a))=U_t\;\pi(a)\;
U_t^{-1}$ and $U_t\Psi=\Psi$ where $U_t=\exp(it H)$. However, $\Psi\not=1$
and $1$ is not normalisable so this is not exactly
parallel to the discussion just performed. To do so,  
consider the state $\omega(.):=<\Omega,\pi(.)\Omega>_{{\cal H}}$. It is not
difficult to see that it produces the unitarily equivalent 
GNS data ${\cal H}_\omega=L_2(\mathbb{R},|\Psi|^2\;dx)$,
$\pi_\omega=\Omega^{-1}\;\pi\;\Omega$ and $\Omega_\omega=1$. Here 
the operator $\Omega$ is defined by $(\Omega\psi)(x)=\Omega(x)\psi(x)$
and its pointwise inverse (as a function) exists a.e. because of the 
separating property. By definition, the unitary operator $W_t$ in this 
GNS representation is now defined as
\be \label{3.46a}
W_t\cdot \pi_\omega(a):=\pi_\omega(\alpha_t(a))=
\Omega^{-1}\pi(\alpha_t(a))\Omega=\exp(itH_\omega)\pi_\omega(a)
\exp(-itH_\omega)
\ee
hence it corresponds also to an inner automorphism $U^\omega_t$ generated by 
$H_\omega=\Omega^{-1}H\Omega$. Obviously, $H_\omega$ annihilates 
$\Omega_\omega=1$ so that in fact $W_t=U^\omega_t$. 

This is also the reason why in the embedding sector the action was not by
conjugation but by action from the left: 
Here the would be unitary operator 
$\exp(i P_\pm(I') X_\pm(\partial I))$ or better its 
rigorous replacement (\ref{3.29}) does leave the GNS vacuum 
$\Omega=1$ invariant
because we know from \cite{LOST} that the GNS representation can be identified
with a space of square integrable functions of generalised ``connections'' 
$P_\pm(x)$, the 
unit operator is represented as the vector equal to unity while $X_\pm(x)$
acts by functional derivation by $P_\pm(x)$. Thus, (\ref{3.29}) applied to 
$1$ yields $1$ again. For the matter sector we have no such result at our 
disposal and it is in fact also not necessary.

To summarise this discussion, the fact that our quantisation of 
$(\exp(\pm iY_\pm(I')^2/4)$ acts by conjugation on the GNS Hilbert space
is not at all surprising but rather a priori a generic feature of the 
action of a Hamiltonian flow in an invariant GNS representation and only
in special cases can one replace the conjugation by a simple action from 
the left. Notice that $D_\pm[f]=D^E_\pm[f] \pm D^M_\pm[f]$ and that in fact
$\alpha^\pm_\varphi=\alpha^{E,\pm}_\varphi\otimes\alpha^{M,\pm}_\varphi$
since matter and embedding variables commute. Therefore a possible 
quantisation of the Hamiltonian constraint resulting from these 
considerations and (\ref{3.22}) is given by
\ba \label{3.47}
&& \tilde{C}[f]\;|\gamma_+,k_+,l_+>\otimes |\gamma_-,k_-,l_->
\nonumber\\
&=& [\sum_{v\in V(\gamma_+)}\; f(v)\;
\{[\exp(iP_+(I'(\gamma_+,v)) X_+(\partial I(\gamma_+,v)))]^\wedge
\otimes
[\exp(iY_+(I'(\gamma_+,v))^2/4)]^\wedge-{\rm id}_{{\cal H}_+}]
\otimes {\rm id}_{{\cal H}_-}\}\; Q(v)]
\cdot\nonumber\\ &&
|\gamma_+,k_+,l_+>\otimes |\gamma_-,k_-,l_->
\nonumber\\
&& 
-
[\sum_{v\in V(\gamma_-)}\; f(v)\;
{\rm id}_{{\cal H}_+}\otimes
\{[\exp(iP_-(I'(\gamma_-,v)) X_-(\partial I(\gamma_-,v)))]^\wedge\otimes
[\exp(-iY_-(I'(\gamma_-,v))^2/4)]^\wedge-{\rm id}_{{\cal H}_-}]\}\; Q(v)]
\cdot\nonumber\\ &&
|\gamma_+,k_+,l_+>\otimes |\gamma_-,k_-,l_->
\ea
Here $(\gamma,v)\mapsto I(\gamma,v),\; (\gamma,v)\mapsto I'(\gamma,v)$ 
are choices of intervals, for each graph $\gamma$ and 
vertex $v\in V(\gamma)$ with the following properties:\\
1. $I'(\gamma,v)=[v,v']$ where $v'$ lies in between $v$ and the next neighbour
of $v$ (counted in the direction of the chosen orientation of $S^1$, i.e.
to the right of $v$). \\
2. $I(\gamma,v)=[v_l,v_r]$ where $v\in [v_l,v_r]$ and $v_l,v_r$ lie in 
between $v$ and both of its next neighbours. \\
The operators displayed are given explicitly in (\ref{3.29}), (\ref{3.41})
and (\ref{3.42}). Let $\varphi_{\gamma,v}$ be any diffeomorphism of $S^1$ 
with the property that $\varphi_{\gamma,v}(v)=v'$ and that any other vertex
of $\gamma$ is left invariant. Then by construction, see (\ref{3.25})
\ba \label{3.48}
&& \tilde{C}[f]\;|\gamma_+,k_+,l_+>\otimes |\gamma_-,k_-,l_->
\nonumber\\
&=& [\sum_{v\in V(\gamma_+)}\; f(v)\;
\{(T_{\varphi_{\gamma_+,v}(\gamma_+),k_+}\; T_{\gamma_+,-k_+}\otimes
W_{\varphi_{\gamma_+,v}(\gamma_+),l_+}\; W_{\gamma_+,-l_+}
-{\rm id}_{{\cal H}_+}
\otimes {\rm id}_{{\cal H}_-}\}\;Q(v)]
\times\nonumber\\&&
|\gamma_+,k_+,l_+>\otimes |\gamma_-,k_-,l_->
\nonumber\\
&& 
-
[\sum_{v\in V(\gamma_-)}\; f(v)\;
\{{\rm id}_{{\cal H}_+}\otimes
(T_{\varphi_{\gamma_-,v}(\gamma_-),k_-}\; T_{\gamma_-,-k_-}\otimes
W_{\varphi_{\gamma_-,v}(\gamma_-),l_+}\; W_{\gamma_-,-l_-}
-{\rm id}_{{\cal H}_-})\}]\; Q(v)
\times\nonumber\\&&
|\gamma_+,k_+,l_+>\otimes |\gamma_-,k_-,l_->
\nonumber\\
&=& 
\sum_{v\in V(\gamma_+)}\; f(v)\;\lambda(v)\;
\{|\varphi_{\gamma_+,v}(\gamma_+),k_+,l_+>
-|\gamma_+,k_+,l_+>\} \otimes |\gamma_-,k_-,l_->
\nonumber\\
&& 
-
\sum_{v\in V(\gamma_-)}\; f(v)\;\lambda(v)
|\gamma_+,k_+,l_+>\otimes \;
\{|\varphi_{\gamma_-,v}(\gamma_-),k_-,l_->
-|\gamma_-,k_-,l_->\} 
\nonumber\\
&=&
[\sum_{v\in V(\gamma_+)}\; f(v)\;
\{U(\varphi_{\gamma_+,v},{\rm id}_{{\rm Diff}(S^1)})-{\rm id}_{\cal H}\}]
\;Q(v)
\;\;\;|\gamma_+,k_+,l_+>\otimes |\gamma_-,k_-,l_->
\nonumber\\
&& -[\sum_{v\in V(\gamma_-)}\; f(v)\;
\{U({\rm id}_{{\rm Diff}(S^1)},\varphi_{\gamma_-,v})-{\rm id}_{{\cal H}}]
\;Q(v)
\;\;\;|\gamma_+,k_+,l_+>\otimes |\gamma_-,k_-,l_->
\ea
where $\lambda(v)$ is the eigenvalue of $Q(v)$ on 
$|\gamma_+,k_+,l_+>\otimes |\gamma_-,k_-,l_->$. Obviously, 
\be \label{3.49}
l[C[f]a]=0\;\;\forall\;\;f,\;\;a\in \mathfrak{A}
\ee
for any linear functional on $\mathfrak{A}$ satisfying (\ref{2.29}). 
On the other hand, if one would expand the exponentials in (\ref{3.47}) 
to linear order (which formally would become more and more exact the closer 
$\varphi_{\gamma,v}$ is to the identity) we would obtain (\ref{3.42}) which
was derived directly from the classical expression. We therefore managed 
to find a proper quantisation of $C[f]$ on the kinematical Hilbert space
by methods developed for LQG 
whose kernel includes the solutions of (\ref{2.29}).

\section{Discussion}
\label{s4}

The most interesting question is of course what aspects of the new 
quantisation techniques developed for the PFT model are likely to be extendable
to the 4D LQG situation in the absence of the luxury of having an exact 
solution. Since of course we cannot answer this question, the following
list can only be speculative in nature:
\begin{itemize}
\item[I.] {\it Algebraic Structure}\\
For the PFT model it would have been impossible to match the solution spaces 
of $D_+,D_-$ and $D,C$ respectively if one had not kept the $C=D_+-D_-$
term in $\tilde{C}=C/\sqrt{\det(q)}$ intact. What we mean by this is 
that in LQG one considers the combination 
\be \label{4.1}
e_a^j\propto \epsilon_{abc}\epsilon^{jkl} E^b_k E^c_l/\sqrt{\det(q)}\propto
\{A_a^j,V\}
\ee
where $V$ is the LQG volume operator and directly quantises this combination
upon replacing the connection $A$ by a holonomy. Here the Euclidian piece 
of the Lorentzian Hamiltonian constraint (which plays an important role
in the quantisation of the Lorentzian constraint) is given by 
\be \label{4.1a}
\tilde{C}=C/\sqrt{\det(q)},\;\;C= 
B^a_j\epsilon_{abc}\epsilon^{jkl} E^b_k E^c_l
\ee
where $B$ is the magnetic field of $A$. The classical kernel is determined
entirely by the density two object $C$. Analogously, in the PFT model we 
could also 
have quantised the $P_\pm X'_\pm/\sqrt{\det(q)}$ term differently, e.g.
based on the identity
\be \label{4.2}
P_\pm(x) X'_\pm(x)/\sqrt{\det(q)}(x)\propto
P_\pm(x) \sqrt{-X'_\pm(x)/X'_\mp(x)}=
2P_\pm(x)\{V(I),P_\mp(J)\}
\ee
where $I$ is an arbitrarily short interval containing $x$ and $J$ overlaps 
only the end point of $I$. The integral over (\ref{4.2}) would then become 
well defined in an triangulation regularisation. However, it would 
not allow to make contact with the $D_\pm$ quantisation. Similarly, in 4D LQG
it could be desirable to leave $C$ intact and to quantise $1/\sqrt{\det(q)}$
as a factor similar as for PFT. This has already been done as a part of the 
quantisation of scalar field contribution to the Hamiltonian \cite{QSD}.
\item[II.] {\it Role of Diffeomorphisms}\\
In the PFT model it proved convenient to parametrise the ambiguities in the 
loop attachment, here the choices of the intervals $I(\gamma,v),I'(\gamma,v)$,
in terms of spatial diffeomorphisms of compact support about $v$. In 4D LQG
one could proceed analogously. As a warm -- up, a quantisation of the 
Husain -- Kucha{\v r} \cite{HK} model along those lines, whose complete 
solution by LQG techniques is well known, is now being completed \cite{TT}.
That is to say, as already observed in \cite{QSD}, while the generator 
of infinitesimal differmorphisms $D_a,\;a=1,2,3$ does not exist in 4D LQG
due to the discontinuity of its one parameter subgroups, what can be defined 
is the operator corresponding to $C_j=E^a_j C_a/\sqrt{\det(q)}$ because 
it is scalar density of weight one \cite{QSD} rather than a covector 
density of weight one as $C_a$. This is analogous to quantising in PFT 
$\tilde{D}=D/\sqrt{\det(q)}$ (which exists, see below) rather than $D$ 
(which does not exist, see above). This object was already quantised using 
the ordinary LQG techniques in \cite{master,AQG} which puts the quantisations
of the spatial diffeomorphism constraints and the Hamiltonian constraints 
on equal footing\footnote{This is sometimes spelled out as a cricism: Why 
should the infinitesimal Hamiltonian constraint exist while only finite
spatial diffeomorphisms exist?}. However, its consistency with the 
Husain -- Kuch{\v r} quantisation was not yet verified which will be 
the subject of \cite{TT}.
\item[III.] {\it Constraint Algebra}\\
Let us now consider the hypersurface deformation algebra. In section 3 
we quantised the spatial diffeomorphism constraint as finite diffeomorphisms
as is customary in LQG. However, in order to see whether the quantisation
of the Hamiltonian constraint reproduces the algebra $\mathfrak{H}$ at the 
quantum level we need the infinitesimal generator $D$ which however does not 
exist. As just mentioned, the object that does exist is 
$\tilde{D}=D/\sqrt{\det(q)}$. Thus, we consider the classically equivalent
algebra of $\tilde{C},\tilde{D}$. Since $C=D_+ -D_-$ and $D=D_+ + D_-$, 
in view of (\ref{3.48}) the quantisation of $\tilde{D}$ is immediate and 
differs only by a sign from (\ref{3.48}), that is
\ba \label{4.3}
&& \tilde{D}[f]\;|\gamma_+,k_+,l_+>\otimes |\gamma_-,k_-,l_->
\nonumber\\
&=&
[\sum_{v\in V(\gamma_+)}\; f(v)\;
\{U(\varphi_{\gamma_+,v},{\rm id}_{{\rm Diff}(S^1)})-{\rm id}_{\cal H}\}]
\;Q(v)
\;\;\;|\gamma_+,k_+,l_+>\otimes |\gamma_-,k_-,l_->
\nonumber\\
&& +[\sum_{v\in V(\gamma_-)}\; f(v)\;
\{U({\rm id}_{{\rm Diff}(S^1)},\varphi_{\gamma_-,v})-{\rm id}_{\cal H}\}]
\;Q(v)
\;\;\;|\gamma_+,k_+,l_+>\otimes |\gamma_-,k_-,l_->
\nonumber\\
\ea
Equations (\ref{3.38}) and (\ref{4.3}) define $\tilde{D}_\pm[f]$ via 
\be \label{4.4}
\tilde{C}[f]=:\tilde{D}_+[f]-\tilde{D}_-[f],\;\;
\tilde{D}[f]=:\tilde{D}_+[f]+\tilde{D}_-[f]
\ee
It is of course immediate that the dual of the 
commutator between the $\tilde{D}_\pm[f]$ still annihilates the kernel 
of $D_\pm$ defined in (\ref{2.29}). But we want to see whether the 
algebra of the classical $\tilde{D}_\pm$ is reproduced and not only 
the kernel. The notationally somewhat tedious calculation is carried out 
in appendix \ref{sa}. The result is as follows:\\
For simplicity and equivalently, we compare the algebra of the operators 
$\tilde{D}_\pm$ with their classical counterpart. The classical Poisson
algebra of the $\tilde{D}_\pm$ closes but only with the structure functions.
We show, that there does exist a quantisation of the right hand side of that
classical Poisson algebra, following the techniques described in section 
\ref{s3}, such that the quantum commutators are {\it precisely} reproduced.
By this we mean that similar as in \cite{QSD} the right hand side of the 
Poisson bracket between Hamiltonian constraints can be quantised by the 
same techniques as for the Hamiltonian constraint itself, in particular
it has to be written in terms of the finite diffeomorphism approximation
to the classical infinite generators. One may call this a soft anomaly in the 
sense that while it is not possible to reproduce the algebra $\mathfrak{H}$
at the quantum level in its infinitesimal version, there is a substitute 
to it which 1. can be seen precisely as the quantisation of $\mathfrak{H}$ 
in terms of finite diffeormorphisms, 2. closes with the correct factor
ordering so that solutions of the constraints are not subject to any
anomalous extra conditions and 3. yields the correct structure functions
in some deformation again caused by the discontinuity of the representation.
Such anomlaies are not troublesome as they do not lead to a mismatch between
classical and quantum physical degrees of freedom. 
\item[IV.] {\it Other Models}\\
Such soft anomalies 
were also recently observed in 2+1 gravity with a cosmological constant 
quantised \'a la LQG \cite{Perez}. In more detail, in \cite{Perez} the 
authors argue that
the anomaly they found presents an obstruction to quantise 2+1 gravity
with LQG methods as in \cite{QSD}. We would like to clarify 
this statement as follows:\\
The constraints of 2+1 gravity with nonvanishing cosmological constant
$\Lambda$ are the Gauss constraint $G$ and the curvature constraint 
$C=F+\Lambda E$ where $F$ is the curvature of the connection and $E$ is 
the volume 2-form built from its conjugate momentum. Classically we have 
(symbolically) $\{G,G\}=G,\;\{G,C\}=C,\;\{C,C\}=G$ with structure constants.
There is no quantisation ambiguity as far as $G$ is concerned but $C$ involves
the choice of a loop attachment just as in 3+1 GR. The commutator algebra
yields (symbolically) $[G,G]=G,\;[G,C]=C,\;[C,C]={\rm Tr}(h)\;G$ where 
$h$ is the holonomy of a loop depending on the choice of loop attachment
(see formula (40) in \cite{Perez} and below).
Could one shrink the loop to a point then we could replace the trace 
by a constant, however, this is not possible due to the discontinuity
of the representation. The point is now that ${\rm Tr}(h)$ is gauge invariant.
Therefore one can also write $[C,C]=G\; {\rm Tr}(h)$, hence {\it the Gauss
constraint correctly appears to the left of the anomalous structure 
functions}. 
Accordingly, linear functionals satisfying $l[G T]=l[C T]=0$ for all 
spin network functions $T$ and all $G,C$ are not subject to any extra 
conditions coming from the above soft anomaly. Hence, there is no 
inconsistency. What does not work is to apply group averaging techniques
because of the anomalous structure functions. Thus, one has to rely 
on alternative techniques such as the master constraint \cite{master}
or one has to construct the space of solutions by hand and equip it
with an inner product wrt which the $^\ast-$algebra of observables 
is faithfully represented. As the PFT example reveals
where something similar happens, this can in principle be done 
and the presence of the anomalous structure functions does not at all imply 
that the LQG quantisation techniques fail.
\item[IV.] {\it Quantisation Ambiguities}\\
The quantisation of the 4D LQG Hamiltonian constraint suffers from 
various quantisation ambiguities. It often argued, that these ambiguities 
might be drastically reduced once one manages to reproduce the hypersurface
deformation algebra at the quantum level. One can only hope to be able to 
do this in terms of finite diffeomorphisms as in the PFT model due to 
the discontinuity of the representation, that is, one has to be ready to accept
a soft anomaly. Now as far as the PFT model is concerned, some aspects of the 
ambiguities of the Wilson like quantisation of the Hamiltonian constraint 
are indeed improved by making use of the $D_\pm$ reformulation. More in 
detail, no factor ordering or representation choice ambiguities in the 
quantisation of the would be $\exp(iD_\pm[f])$ arise, that part of the 
quantisation is clean and unambiguous. The ambiguities arise only in the 
choice of the support of the finite diffeomorphisms and in the representation
choice of the holonomies entering the inverse volume operator. Both 
ambiguities have absolutely no effect on the space of solutions. 
This suggests that in 4D LQG the only worrysome ambiguity lies in the 
choice of the loop attachment. This kind ambiguity becomes worse the higher
the spatial dimension, because it depends on the choice of a vector field
of compact support whose integral curves define this diffeomorphism. In one 
spatial dimension, there is only one direction, but in higher dimensions the 
choice of vector field becomes much less trivial.
We will come back to this point in \cite{TT}.
\item[V.] {\it Kinematical Semiclassical States}\\
Some important tool for answering the question whether a set of operators
qualifies as the quantisation of a given set of classical functions on the 
phase space are semiclassical (minimal uncertainty) states. This is especially
important in 4D LQG where the complicated volume operator prevents us from
doing any analytic calculation so that one has to rely on semiclassical
techniques which allow for suitable approximations \cite{AQG}. In the 
PFT model the volume operator is of course under full analytical control.
However, the kinematical\footnote{The semiclassical states should not be 
solutions to the proposed quantum constraints as otherwise one cannot check
by means of them whether the constraints have been quantised correctly.}
semiclassical states proposed so far \cite{Coherent} are 
not suitable to answer the question whether the Hamiltonian constraint
of 4D LQG has been quantised correctly. The reason for this is as follows:\\
The semiclassical states of usual (free) field theories are coherent 
superpositions of the corresponding Fock basis states. This is not possible
in LQG because, while the spin network states provide a Fock like basis,
in contrast to Fock representations the LQG representation is not separable.
This implies that any semiclassical vector state is sensitive to the excitations 
of an at most countable set of edges. However, the Hamiltonian constraint 
of 4D LQG \cite{QSD} has the peculiar property of always modifying the 
graph of the state on which it acts, no matter how large it is. As a result,
the expectation value of the Hamiltonian constraint in such vector states
is always zero even if they ascribe the correct holonomy and flux expectation 
values to the edges (and dual faces) of the graph on which the vector state 
depends. In the PFT model, we can ask a similar question, namely whether 
{\it kinematical} semiclassical vector states exist that probe the correctness
of the quantum constraints\footnote{{\it Physical} coherent states
have been constructed explicitly in \cite{Varadarajan} for each of the 
superselection sectors.}. For PFT this is a somewhat academic question 
because the finite $D_\pm$ operators have a geometric action and there is 
no doubt that they have been quantised correctly. However, as a test for 
4D LQG this is an interesting question to ask. Notice that also in PFT
the constraints of the form 
$U(\varphi^+_{\gamma^+,v},1)-1,\;U(1,\varphi^-_{\gamma^-,v},1)-1$ 
always modify the 
graph on which the operator acts, they are never graph preserving. However,
there is a difference to 4D LQG: While the 4D LQG operator changes the number
of vertices and edges of a graph, the PFT operators do not do that
since the finite diffeomorphisms they generate are confined to the circle. 
This simplification might enable one to improve the situation for the 
PFT model whose kinematical Hilbert space is also non separable. One 
idea \cite{books} is to use vector states that are superpositions 
of a given semiclassical state and all its images under the repeated 
action of the Hamiltonain constraint (fractal graph coherent state). 
However, preliminary investigation indicates that this still does not
work. If this looks hopeless in the PFT model, then it is probably also in 
the much more complicated 4D LQG theory and possibly then the question 
of the correctness of the quantisation can only be answered at the physical
Hilbert space level.
\item[VI.] {\it Observables}\\
The observables of the theory have been constructed in all detail in 
\cite{Varadarajan}. Their dual acts on the space of solutions to 
(\ref{2.29}) from which follows that also the dual of their commutator
with $\tilde{C},\tilde{D}$ annihilates physical states. However,
given the fact that 
here we have complementary restrictions on $k_0,l_0$ as compared to 
\cite{Varadarajan}, this induces different restrictions on the 
quantisation of the observables 
\be \label{4.5a}
O^\pm_f:=O_{Y_\pm,f}:=\sum_{n\in \mathbb{Z}}\; \hat{f}_n\; 
O^\pm_{Y_\pm,n}
\ee
where the $\hat{f}_n$ are complex numbers without a priori restriction,
see (\ref{2.17}). We will sketch these in the following paragraph. 

Following Hardy -- Littlewood theory \cite{LH}, there are several means 
of sequences of real or complex numbers $a_m\in \mathbb{C}$ which are 
widely used by physicists and which have the property to 1. produce well
defined limits even if the sequence itself does not converge and 2. to
reproduce the limit of the sequence if it does converege. The {\it
Cesaro mean}
\be \label{4.5b}
C[a]:=\lim_{M\to \infty} \frac{1}{2M-1}
\sum_{m\in \mathbb{Z};\;|m|\le M-1}\; a_m
\ee
used extensively in section is only one of them. To avoid confusion, the 
Cesaro mean appeared {\it naturally} in section \ref{s3} in an exact equality, 
it was not used in order to ``sweep divergencies under the  rug''.

The Cesaro mean also naturally appears in the representation of the 
observables as follows:
Since $Y_\pm[I']$ does not exist, it is natural, following
\cite{Varadarajan}, to try to define instead the exponential
\be \label{4.5c}
\exp(iO^\pm_f) \;|\gamma^\pm,k_\pm,l_\pm>,\;O^\pm_f:=\sum_n \; \hat{f}_n\; 
O^\pm_n 
\ee
By exactly the same 
calculation displayed in \cite{Varadarajan} one arrives at
\be \label{4.5d}
\exp(iO^\pm_f) \;|\gamma^\pm,k_\pm,l_\pm>=
\exp(i\sum_{I\in\gamma^\pm}\; Y_\pm(I)\;l_I)\; |\gamma^\pm,k_\pm,l_\pm>=
\ee
which is well defined provided that
\be \label{4.5e}
l_I:=\sum_n\;\hat{f}_n\;  e^{in\hbar k_I/R} \in l_0 \mathbb{Z} 
\ee
for any $k_I\in k_0\mathbb{Z}$.
Let us write $\hat{f}_n=:l_0\hat{g}_n$ and let $\alpha:=\hbar k_0/R$. Then 
requirement (\ref{4.5d}) simply reads
\be \label{4.5f}
z_m:=\sum_n\;\hat{g}_n\;  e^{i\alpha n m} \in \mathbb{Z} 
\ee
for any $m\in \mathbb{Z}$ as a condition of the Fourier coefficients 
$\hat{g}_n$. Using the Cesaro mean, we can translate this into 
the requirement that the integer valued sequence $(z_m)$ should be such 
that 
\be \label{4.5g}
\hat{g}_n:=C[a^{(n)}],\;a^{(n)}_m:=e^{-i\alpha mn}\; z_m
\ee
converges for any $n\in \mathbb{Z}$. 
Hence, also here we see a reduction in 
the number of possible $f$ similar to \cite{Varadarajan}.
\item[VII.] {\it Methodology}\\
Anybody who has gone through the details of section \ref{s3} will admit
that nobody would ever have thought of the constructions there, if the 
alternative formulation in terms of the $D_\pm$ constraints would not have been 
at our disposal. Therefore one can rightfully ask, why one should go through
these indirect constructions if a simpler solution is available and vice versa
what use these constructions have, if a simpler solution is not available. 
Perhaps a possible answer could be to look for classical reformulations 
of a constraint algebra in terms of simpler algebras or for algebras
of elementary observables that are more adapted to the dynamics 
of the theory. This is already the case in free, massive scalar
field theories: There one writes the Hamiltonian in terms of the natural 
creation and annihilation
operators induced by the Hamiltonian and not of some completely different 
free field
Hamiltonian (say with a different mass). It is not even possible to define 
the Hamiltonian with a different mass in a Fock representation with a 
given mass, one of the consequences of Haag's theorem \cite{Haag}. 
In LQG the holonomy flux algebra is well adapetd to the 
spatial diffeomorphism constraint but not obviously as far as the 
Hamiltonian constraint is concerned. Maybe the choice of the holonomy
flux algebra as a classical starting point for the quantisation should be 
reconsidered.  On the other hand, at the very least it 
is reaffirming, that a non trivial fraction of the techniques introduced 
in \cite{QSD} turn out to be useful and correct and also in  the completely
solvable PFT.
\end{itemize}

~\\
\\
{\large Acknowledgements}\\
\\
We are very grateful to Madhavan Varadarajan for awakening our interest 
in PFT as test laboratory for LQG during a detailed discussion at 
the 2010 Zakopane LQG workshop. Furthermore, we thank Alok Laddha and 
Madhavan Varadarajan for useful comments on an earlier draft of this 
paper.\\ 
The part of the research performed at the Perimeter Institute for 
Theoretical Physics was supported in part by funds from the Government of  
Canada through NSERC and from the Province of Ontario through MEDT.

\begin{appendix}

\section{Constraint Algebra}
\label{sa}

We compute first the classical algebra of the $\tilde{D}_\pm=D_\pm/
\sqrt{\det(q)}$. Notice that while $\det(q)=-X_+' X_-'$ transforms as 
a scalar density of weight two under $D$, it transforms only as a scalar density
of weight one under $D_\pm$. Let $f,g$ be some test 
functions, then from (\ref{2.15})
\ba \label{a.1}
&&
\{\tilde{D}_\pm[f],\tilde{D}_\pm[g]\} =
\int\; dx\int\; dy \; f(x)\; g(y)\;
\{\tilde{D}_\pm(x),\tilde{D}_\pm(y)\}
\nonumber\\
&=&
\int\; dx\int\; dy \; f(x)\; g(y)\;[
\frac{1}{\sqrt{\det(q)}(x)\sqrt{\det(q)}(y)}\;\{D_\pm(x),D_\pm(y)\}
+\frac{D_\pm(x)}{\sqrt{\det(q)}(y)}\;\{\frac{1}{\sqrt{\det(q)}(x)},D_\pm(y)\}
\nonumber\\ &&
-\frac{D_\pm(y)}{\sqrt{\det(q)}(x)}\;\{\frac{1}{\sqrt{\det(q)}(y)},D_\pm(x)\}]
\nonumber\\
&=&
\{D_\pm[\tilde{f}],D_\pm[\tilde{g}]\}
-\int\;dx\; \frac{f(x) D_\pm(x)}{\sqrt{\mp X'_\mp}(x)}
\{D_\pm[\tilde{g}],\frac{1}{\sqrt{\pm X'_\pm(x)}}
+\int\;dy\; \frac{g(y) D_\pm(y)}{\sqrt{\mp X'_\mp}(y)}
\{D_\pm[\tilde{f}],\frac{1}{\sqrt{\pm X'_\pm(y)}}
\nonumber\\
&=&
D_\pm[[\tilde{f}],\tilde{g}]]
-\int\;dx\; \frac{f D_\pm}{\sqrt{\mp X'_\mp}}
[\tilde{g}(\frac{1}{\sqrt{\pm X'_\pm}})'
-\frac{1}{2}\tilde{g}'\frac{1}{\sqrt{\pm X'_\pm}}]
+\int\;dx\; \frac{g D_\pm}{\sqrt{\mp X'_\mp}}
[\tilde{f} (\frac{1}{\sqrt{\pm X'_\pm}})'
-\frac{1}{2} \tilde{f}' \frac{1}{\sqrt{\pm X'_\pm}}]
\nonumber\\
&=&
D_\pm[[\tilde{f}],\tilde{g}]]
-\frac{1}{2}\int\;dx\; \sqrt{\pm X_\pm'}\; D_\pm
[\tilde{f} \tilde{g}'\frac{1}{\sqrt{\pm X'_\pm}}]
[\tilde{g} \tilde{f}' \frac{1}{\sqrt{\pm X'_\pm}}]
\nonumber\\
&=& \frac{1}{2} D_\pm[[\tilde{f}],\tilde{g}]]
= \frac{1}{2} \tilde{D}_\pm[\sqrt{\det(q)}[\tilde{f}],\tilde{g}]]
\nonumber\\
&=& \frac{1}{2} \tilde{D}_\pm[\frac{1}{\sqrt{\det(q)}}[f,g]]
\ea
with $\tilde{f}=f/\sqrt{\det(q)},\tilde{f}=f/\sqrt{\det(q)}$ and one computes 
the Poisson brackets displayed as if $\tilde{f},\tilde{g}$ were independent 
of the phase space. Similarly,
\ba \label{a.2}
&&
\{\tilde{D}_\pm[f],\tilde{D}_\mp[g]\} =
\int\; dx\int\; dy \; f(x)\; g(y)\;
\{\tilde{D}_\pm(x),\tilde{D}_\mp(y)\}
\nonumber\\
&=&
\int\; dx\int\; dy \; f(x)\; g(y)\;[
\frac{1}{\sqrt{\det(q)}(x)\sqrt{\det(q)}(y)}\;\{D_\pm(x),D_\mp(y)\}
+\frac{D_\pm(x)}{\sqrt{\det(q)}(y)}\;\{\frac{1}{\sqrt{\det(q)}(x)},D_\mp(y)\}
\nonumber\\ &&
-\frac{D_\mp(y)}{\sqrt{\det(q)}(x)}\;\{\frac{1}{\sqrt{\det(q)}(y)},D_\pm(x)\}]
\nonumber\\
&=&
\{D_\pm[\tilde{f}],D_\mp[\tilde{g}]\}
-\int\;dx\; \frac{f(x) D_\pm(x)}{\sqrt{\pm X'_\pm}(x)}
\{D_\mp[\tilde{g}],\frac{1}{\sqrt{\mp X'_\mp(x)}}
+\int\;dy\; \frac{g(y) D_\mp(y)}{\sqrt{\mp X'_\mp}(y)}
\{D_\pm[\tilde{f}],\frac{1}{\sqrt{\pm X'_\pm(y)}}
\nonumber\\
&=&
-\int\;dx\; \frac{f D_\pm}{\sqrt{\pm X'_\pm}}
[\tilde{g}(\frac{1}{\sqrt{\mp X'_\mp}})'
-\frac{1}{2}\tilde{g}'\frac{1}{\sqrt{\mp X'_\mp}}]
+\int\;dx\; \frac{g D_\mp}{\sqrt{\mp X'_\mp}}
[\tilde{f} (\frac{1}{\sqrt{\pm X'_\pm}})'
-\frac{1}{2} \tilde{f}' \frac{1}{\sqrt{\pm X'_\pm}}]
\nonumber\\
&=&
-\int\;dx\; f \tilde{D}_\pm \sqrt{\mp X'_\mp}
[\tilde{g}(\frac{1}{\sqrt{\mp X'_\mp}})'
-\frac{1}{2}\tilde{g}'\frac{1}{\sqrt{\mp X'_\mp}}]
+\int\;dx\; g \tilde{D}_\mp \sqrt{\pm X'_\pm}
[\tilde{f} (\frac{1}{\sqrt{\pm X'_\pm}})'
-\frac{1}{2} \tilde{f}' \frac{1}{\sqrt{\pm X'_\pm}}]
\nonumber\\
&=&
\frac{1}{2}\int\;dx\; f \tilde{D}_\pm \sqrt{\mp X'_\mp}
[\tilde{g}\frac{\mp X^{\prime\prime}_\mp}{\sqrt{\mp X'_\mp}^3}
+\tilde{g}'\frac{1}{\sqrt{\mp X'_\mp}}]
-\frac{1}{2}\int\;dx\; g \tilde{D}_\mp \sqrt{\pm X'_\pm}
[\tilde{f} \frac{\pm X^{\prime\prime}_\pm}{\sqrt{\pm X'_\pm}^3}
+\tilde{f}' \frac{1}{\sqrt{\pm X'_\pm}}]
\nonumber\\
&=&
\frac{1}{2}\int\;dx\; f \frac{\tilde{D}_\pm}{\mp X'_\mp}
[\tilde{g} (\mp X'_\mp)]'
-\frac{1}{2}\int\;dx\; g \frac{\tilde{D}_\mp}{\pm X'_\pm}
[\tilde{f} (\pm X'_\pm)]'
\nonumber\\
&=&
-\frac{1}{2}\int\;dx\; f\; D_\pm\; \frac{X'_\pm}{\sqrt{\det(q)}^3}
[g \frac{X'_\mp}{\sqrt{\det(q)}}]'
+\frac{1}{2}\int\;dx\; g\; D_\mp\; \frac{X'_\mp}{\sqrt{\det(q)}^3}
[f \sqrt{X'_\pm}{\sqrt{\det(q)}}]'
\ea
with no further simplification possible. The modifications in the algebra 
of the $\tilde{D}_\pm$ of course arise because of the nontrivial 
transformation behaviour of $\det(q)$ wrt $D_\pm$. As expected, the new 
algebra closes albeit with non trivial structure functions.

We now turn to the quantum computation. For notational simplicity, given 
graphs $\gamma^\pm$ and $v\in V(\varphi^\pm)$ we denote 
$\varphi^\pm_v:=\varphi_{\gamma^\pm,v}$ and 
$\gamma^\pm_v:=\varphi^\pm_v(\gamma^\pm)$ where $(\gamma,v)\mapsto 
\varphi_{\gamma,v}$ is the choice of diffeomorphism depending on $\gamma,\;
v\in V(\gamma)$ and specified in section \ref{s3}. Likewise, for 
$\tilde{v}\in V(\gamma^\pm_v)$ we denote $\varphi^\pm_{v\tilde{v}}:=
\varphi_{\gamma^\pm_v,\tilde{v}}$ and $\gamma^\pm_{v\tilde{v}}:=
\varphi^\pm_{v\tilde{v}}(\gamma^\pm_v)$. Finally, we define for a given 
charge network $|\gamma^+,k_+,l_+>\otimes|\gamma^-,k_-,l_->$ by 
$\lambda(\gamma^+,\gamma^-,v)$ the eigenvalue of $Q(v)$ on this eigenvector. 
As follows from
(\ref{3.20}), (\ref{3.21}) and (\ref{3.22}), these eigenvalues depend 
only on those $k^\pm_I, I\in \gamma^\pm$ such that $v\in \partial I$.
In particular it vanishes if $v$ is not a common vertex of both $\gamma^+$ 
and $\gamma^-$. Moreover, these eigenvalues are diffeomorphism invariant 
in the sense that 
\be \label{a.3}
Q(v)|\gamma^+,k_+,l_+>\otimes|\gamma^-,k_-,l_->
=U(\varphi^{-1},\varphi^{-1}) Q(\varphi(v)) 
|\varphi(\gamma_+),k_+,l_+>\otimes  |\varphi(\gamma_-),k_-,l_->
\ee

With these preparations we compute first the $\tilde{D}_+,\tilde{D}_+$
commutator (the $\tilde{D}_-,\tilde{D}_-$ commutator is completeley analogous).
We display only the graph dependence
for notational simplicity since the charges are carried with the vertices 
under diffeomorphisms and are not changed. Also the operator $Q(v)$ 
does not change the charges. We begin with
\ba \label{a.4}
&& [\tilde{D}_+[f],\tilde{D}_+[g]]\; |\gamma^+,\gamma^->
\nonumber\\
&=& 
\sum_{v\in V(\gamma^+)}\; g(v)\; \lambda(\gamma^+,\gamma^-,v)\;
D_+[f]\;[|\gamma^+_v,\gamma^-> -|\gamma^+,\gamma^->]
\nonumber\\
&& - 
\sum_{v\in V(\gamma^+)}\; f(v)\; \lambda(\gamma^+,\gamma^-,v)\;
D_+[g]\;[|\gamma^+_v,\gamma^-> -|\gamma^+,\gamma^->]
\nonumber\\
&=& 
\sum_{v\in V(\gamma^+)}\; g(v)\; \lambda(\gamma^+,\gamma^-,v)\;
\{\sum_{\tilde{v}\in V(\gamma^+_v)} \; f(\tilde{v})\; 
\lambda(\gamma^+_v,\gamma^-,\tilde{v})
[|\gamma^+_{v\tilde{v}},\gamma^->-|\gamma^+_v,\gamma^->]
\nonumber\\
&&
-\sum_{\tilde{v}\in V(\gamma^+)} \; f(\tilde{v})\; 
\lambda(\gamma^+,\gamma^-,\tilde{v})
[|\gamma^+_{\tilde{v}},\gamma^->-|\gamma^+,\gamma^->]\}
\nonumber\\
&&
-\sum_{v\in V(\gamma^+)}\; f(v)\; \lambda(\gamma^+,\gamma^-,v)\;
\{\sum_{\tilde{v}\in V(\gamma^+_v)} \; g(\tilde{v})\; 
\lambda(\gamma^+_v,\gamma^-,\tilde{v})
[|\gamma^+_{v\tilde{v}},\gamma^->-|\gamma^+_v,\gamma^->]
\nonumber\\
&&
-\sum_{\tilde{v}\in V(\gamma^+)} \; g(\tilde{v})\; 
\lambda(\gamma^+,\gamma^-,\tilde{v})
[|\gamma^+_{\tilde{v}},\gamma^->-|\gamma^+,\gamma^->]\}
\ea
Consider the terms proportional to $|\gamma^+,\gamma^->$ which is independent 
of $v,\tilde{v}$. These involve a 
sum over $v,\tilde{v}\in V(\gamma^+)$ with coefficient 
\be \label{a.5}
g(v)\lambda(\gamma^+,\gamma^-,v) 
f(\tilde{v})\lambda(\gamma^+,\gamma^-,\tilde{v}) 
-g(v)\lambda(\gamma^+,\gamma^-,v) 
f(\tilde{v})\lambda(\gamma^+,\gamma^-,\tilde{v}) 
\ee
which is antisymmetric under $v\leftrightarrow \tilde{v}$ and thus 
the corresponding sum vanishes. 

Next consider the terms proportional to $|\gamma^+_v,\gamma^->$ and
$|\gamma^+_{\tilde{v}},\gamma^->$. The former terms involve a sum over 
$v\in V(\gamma^+),\tilde{v}\in V(\gamma^+_v)$ the latter over
$v,\tilde{v} \in V(\gamma^+)$. Relabel $v\leftrightarrow \tilde{v}$ in the 
latter sum. Then these two contributions can be written as 
\ba \label{a.6} 
&&
-\sum_{v\in V(\gamma^+),\tilde{v}\in V(\gamma^+_v)}\; 
\lambda(\gamma^+,\gamma^-,v)\;\lambda(\gamma^+_v,\gamma^-,\tilde{v}) 
[g(v)f(\tilde{v})-f(v)g(\tilde{v})]\;|\gamma^+_v,\gamma^->
\nonumber\\
&&
+\sum_{v,\tilde{v}\in V(\gamma^+)}\; 
\lambda(\gamma^+,\gamma^-,v)\;\lambda(\gamma^+,\gamma^-,\tilde{v}) 
[g(v)f(\tilde{v})-f(v)g(\tilde{v})]\;|\gamma^+_v,\gamma^->
\ea
Notice that $V(\gamma^+_v)=[V(\gamma^+)-\{v\}]\cup \{\varphi^+_v(v)\}$
We split the sum in the first term in (\ref{a.6}) into $\tilde{v}\not=v$ and 
$\tilde{v}=\varphi^+_v(v)=:\hat{v}$. For $\tilde{v}\not=\hat{v}$ we have 
$\lambda(\gamma^+_v,\gamma^-,\tilde{v})=
\lambda(\gamma^+,\gamma^-,\tilde{v})$ because the vertices in $\gamma^+$ 
different from $v$ are not moved by $\varphi^+_v$ and the eigenvalue 
$\lambda(\gamma^+,\gamma_-,v)$ depends only on the infinitesimal neighbourhood
of $v$. It follows that the terms $\tilde{v}\not=v$ in the first sum and 
those with $\tilde{v}\not=v$ in the second sum in (\ref{a.6}) 
cancel each other and what remains is (notice that the contribution to the 
second sum from $\tilde{v}=v$ vanishes trivially) 
\be \label{a.7} 
-\sum_{v\in V(\gamma^+)}\; 
\lambda(\gamma^+,\gamma^-,v)\;\lambda(\gamma^+_v,\gamma^-,\hat{v}) 
[g(v)f(\hat{v})-f(v)g(\hat{v})]\;|\gamma^+_v,\gamma^->
\ee
Finally consider the terms proportional to $|\gamma_{v\tilde{v}},\gamma^->$
whose contribution is given by
\be \label{a.8}
\sum_{v\in V(\gamma^+),\;\tilde{v}\in V(\gamma^+_v)} 
\lambda(\gamma^+,\gamma^-,v)\;
\lambda(\gamma^+_v,\gamma^-,\tilde{v})
[g(v)\; f(\tilde{v})-f(v)\; g(\tilde{v})]\; 
[|\gamma^+_{v\tilde{v}},\gamma^->
\ee
Again we split the sum over $\tilde{v}$ into $\tilde{v}\in V(\gamma^+)-\{v\}$
and $\tilde{v}=\hat{v}$. Then (\ref{a.8}) becomes  
\ba \label{a.9}
&& 
\sum_{v,\tilde{v}\in V(\gamma^+),\;v\not=\tilde{v}} 
\lambda(\gamma^+,\gamma^-,v)\;
\lambda(\gamma^+_v,\gamma^-,\tilde{v})
[g(v)\; f(\tilde{v})-f(v)\; g(\tilde{v})]\; 
|\gamma^+_{v\tilde{v}},\gamma^->
\nonumber\\
&& +\sum_{v\in V(\gamma^+)} 
\lambda(\gamma^+,\gamma^-,v)\;
\lambda(\gamma^+_v,\gamma^-,\hat{v})
[g(v)\; f(\hat{v})-f(v)\; g(\hat{v})]\; 
[|\gamma^+_{v\hat{v}},\gamma^->
\ea
As already shown, $\lambda(\gamma^+_v,\gamma^-,\tilde{v})=
\lambda(\gamma^+,\gamma^-,\tilde{v})$ for $\tilde{v}\not=v$. Therefore the 
first term in (\ref{a.9}) can be written
\be \label{a.10}
\frac{1}{2} \sum_{v,\tilde{v}\in V(\gamma^+),\;v\not=\tilde{v}} 
\lambda(\gamma^+,\gamma^-,v)\;
\lambda(\gamma^+,\gamma^-,\tilde{v})
[g(v)\; f(\tilde{v})-f(v)\; g(\tilde{v})]\; 
[|\gamma^+_{v\tilde{v}},\gamma^->-
|\gamma^+_{\tilde{v}v},\gamma^->]
\ee
where we added the same sum relabelled by 
$v\leftrightarrow \tilde{v}$ and divided by $2$. Now notice that by definition
of $\varphi^+_{\gamma,v}$ this diffeomorphism only notices the next neighbour 
structure of $\gamma^+$. Therefore, $\varphi_{v\tilde{v}}=
\varphi_{\tilde{v}v}$ unless $\tilde{v}\in\{l(v),r(v)\}$ is a left or right
next neighbour of 
$v$ in $\gamma^+$. Consequently, 
also $\gamma^+_{v\tilde{v}}=\gamma^+_{\tilde{v}v}$ then. 
Thus what remains from (\ref{a.10}) is 
\be \label{a.11}
\frac{1}{2} \sum_{v\in V(\gamma^+),\;\tilde{v}\in\{l(v),r(v)\}} 
\lambda(\gamma^+,\gamma^-,v)\;
\lambda(\gamma^+,\gamma^-,\tilde{v})
[g(v)\; f(\tilde{v})-f(v)\; g(\tilde{v})]\; 
[|\gamma^+_{v\tilde{v}},\gamma^->-
|\gamma^+_{\tilde{v}v},\gamma^->]
\ee
Combining (\ref{a.7}), (\ref{a.9}) and (\ref{a.11}) we find 
\ba \label{a.12}
&& [\tilde{D}_+[f],\tilde{D}_+[g]]\; |\gamma^+,\gamma^->
\nonumber\\
&=&
\frac{1}{2} \sum_{v\in V(\gamma^+),\;\tilde{v}\in\{l(v),r(v)\}} 
\lambda(\gamma^+,\gamma^-,v)\;
\lambda(\gamma^+,\gamma^-,\tilde{v})
[g(v)\; f(\tilde{v})-f(v)\; g(\tilde{v})]\; 
[|\gamma^+_{v\tilde{v}},\gamma^->-
|\gamma^+_{\tilde{v}v},\gamma^->]
\nonumber\\
&&+ \sum_{v\in V(\gamma^+)} 
\lambda(\gamma^+,\gamma^-,v)\;
\lambda(\gamma^+_v,\gamma^-,\hat{v})
[g(v)\; f(\hat{v})-f(v)\; g(\hat{v})]\; 
[|\gamma^+_{v\hat{v}},\gamma^->-|\gamma^+_v,\gamma^->]
\ea
Consider the first term in (\ref{a.12}):\\
Notice that 
\be \label{a.13}
\gamma^+_{v\tilde{v}}=\varphi^+_{v\tilde{v}}\circ\varphi_v(\gamma^+),\;
\gamma^+_{\tilde{v}v}=\varphi^+_{\tilde{v}v}\circ
\varphi_{\tilde{v}}(\gamma^+)
\ee
Here $\varphi^+_v\;\varphi^+_{\tilde{v}v}$ only act on $v$ while 
$\varphi^+_{\tilde{v}}\;\varphi^+_{v\tilde{v}}$ only act on $\tilde{v}$
by shifting these vertices to the right without touching the next neighbour
vertex of the graph they act on. However, since e.g. $\varphi^+_v$ acts 
on $\gamma^+$ while $\varphi^+_{v\tilde{v}}$ acts on $\gamma^+_v$, the 
position $\varphi^+_{v\tilde{v}}(\tilde{v})$ maybe different from 
$\varphi_{\tilde{v}}(\tilde{v})$ because the diffeomorphisms depend 
on the next neighbour structure of the graph they act and these are different
in this case. Consequently $\gamma^+_{v\tilde{v}}\not=\gamma^+_{\tilde{v}v}$
in general. This term is similar in structure to the commutator of two
Hamiltonian constraints in 4D LQG. \\
Now consider the second term in (\ref{a.12}):\\
This term arises from a second action of an infinitesimal diffeomorphism 
on the shifted vertex $\hat{v}=\varphi^+_v(v)$. This term is absent in 
4D LQG because due to the properties of the volume operator there (it does
not act on coplanar vertices) it does not act on the vertices it 
creates\footnote{
Here the infinitesimal diffeomorphisms $\varphi^+_v$ could be argued to 
simultaneously create a new vertex $\hat{v}$ and annihilate an old vertex
$v$. Or one could say that the charges on the segment $[v,\hat{v}]$ were changed
from $k^+_{[v,r(v)]},l^+_{[v,r(v)]}$ to 
$k^+_{[l(v),v]},l^+_{[l(v),v]}$. This is similar to 4D LQG but not quite 
analogous because charges are just shifted but never changed on segments
of the graph in question.}. Now it is easy to see, again due to the properties
of the volume operator, that the second term in 
(\ref{a.12}) also vanishes unless by chance the vertex $\hat{v}$ is also
a vertex of $\gamma^-$. It is easy to extend the prescription for the 
$\varphi_{\gamma,v}$ so that this possibility is avoided by hand, 
thus becoming 
a prescription of the form $\varphi_{\gamma^+,v;\gamma^-},\;v\in V(\gamma^+)$ 
and similar for 
$\varphi_{\gamma^-,v;\gamma^+},\;v\in V(\gamma^-)$. Hence with this extended
prescription understood, precisely for the same reason as in 4D LQG
only the first term in (\ref{a.12}) survives because the Hamiltonian 
constraint does not act on the vertices it creates which is ultimately
a property of the volume operator.

Comparing (\ref{a.1}) and (\ref{a.12}) we recognise a similar
structure of the ``structure operators'' and of the structure 
functions. Both terms are propotional to 
an infinitesimal $D_+$ diffeomorphism and the structure functions 
(two factors of $1/\sqrt{\det(q)}$ correctly correspond
to the two 
eigenvalues of the inverse volume. The commutator is manifestly {\it local},
only next neighbour vertices or its infinitesimal diffeomorphic image are 
involved in the sum over vertices of $\gamma^+$. Moreover, the terms
of the form 
\be \label{a.14}
g(v)\; f(\tilde{v})-f(v)\; g(\tilde{v})=
g(v)\; [f(\tilde{v})-f(v)]-f(v)\; [g(\tilde{v})-g(v)]
\ee
qualify as discretisations of the bracket $[f,g](v)$. Even the factors $1/2$
in (\ref{a.1}) and (\ref{a.12}) come out the same.
The question is whether (\ref{a.12}) qualifies as a possible 
quantisation of (\ref{a.1}). To answer this question we write (\ref{a.12})
in the form (dropping the second term as just discussed)
\ba \label{a.15}
&&[\tilde{D}_+[f],\tilde{D}_+[g]]\; |\gamma^+,\gamma^->
\nonumber\\
&=& \frac{1}{2} \sum_{v\in V(\gamma^+),\tilde{v}\in\{l(v),r(v)\}} 
[g(v)\; f(\tilde{v})-f(v)\; g(\tilde{v})]\; 
[U(\varphi^+_{v\tilde{v}}\circ\varphi^+_v,1)-
U(\varphi^+_{\tilde{v}v}\circ\varphi^+_{\tilde{v}},1)]Q(\tilde{v})\;Q(v)
|\gamma^+,\gamma^->\nonumber\\
&=& \frac{1}{2} \sum_{v\in V(\gamma^+)} 
\{
[g\; \delta_r f-f\; \delta_r g)](v)\; 
[U(\varphi^+_{v r(v)}\circ\varphi^+_v,1)-
U(\varphi^+_{r(v)v}\circ\varphi^+_{r(v)},1)]Q(r(v))\;Q(v)
|\gamma^+,\gamma^->
\nonumber\\
&& -
[g\; \delta_l f-f\; \delta_l g)](v)\; 
[U(\varphi^+_{v l(v)}\circ\varphi^+_v,1)-
U(\varphi^+_{l(v)v}\circ\varphi^+_{l(v)},1)]Q(l(v))\;Q(v)
\}
|\gamma^+,\gamma^->
\ea
where we have introduced the right and left graph difference
\be \label{a.16}
[\delta_r f](v)=f(r(v))-f(v),\;\;
[\delta_l f](v)=f(v)-f(l(v))
\ee
However, this is {\it precisely} a possible quantisation of (\ref{a.1}):
The two inverse volume functions would have been ordered to the right
and been replaced by two operators $Q$. There is a freedom whether to 
locate both factors of $Q$ at $v$ or maybe one at $r(v),l(v)$. The quantum
computation decides for the latter possibility, both are equivalent 
in the continuum limit of graphs with a large number of vertices.
The bracket $[f,g]$ would have been 
replaced by the discrete difference between neigbouring vertices and 
again there is a choice between left, right or symmetric derivative. 
The quantum computation decides for a mixture of the two. Finally, 
in the limit of large graphs, the two terms in (\ref{a.15}) at given 
$v$ combine to 
\be \label{a.17}
[f,g](v) 
\{[U(\varphi^+_{v r(v)}\circ\varphi^+_v,1)-
U(\varphi^+_{r(v)v}\circ\varphi^+_{r(v)},1)]\;|r(v)-v|
-[U(\varphi^+_{v l(v)}\circ\varphi^+_v,1)
-U(\varphi^+_{l(v)v}\circ\varphi^+_{l(v)},1)]\;|l(v)-v|\}\;Q(v)^2
\ee
Now by definition $U(\varphi^+_v,1)$ is supposed to be the quantisation 
of the would be operator $\exp(iD_+[I_v])$ where $I_v$ is an interval
containing $v$ corresponding to the support of $\varphi^+_v$. Similarly 
$U(\varphi^+_{v\tilde{v}},1)$ is supposed to be the quantisation 
of $\exp(iD_+[I^v_{\tilde{v}}])$ where $I^v_{\tilde{v}}$ is an interval
containing $\tilde{v}$ corresponding to the support of 
$\varphi^+_{v\tilde{v}}$ which in turn depends on $v$. Now, if the
operators $D_+[I]$ existed then we could expand the curly bracket in 
(\ref{a.17}) to first order in the interval lengths as 
\be \label{a.18}
i(D_+[I_v]+D_+[I^v_{r(v)}]-D_+[I_{r(v)}]-D_+[I^{r(v)}_v]
-D_+[I_v]-D_+[I^v_{l(v)}]+D_+[I_{l(v)}]+D_+[I^{l(v)}_v])
\ee
which would classically correspond to $D_+[J]$ where $J$ is an interval
containing $v$ and of interval length corresponding to absolute value 
of the signed sum of the intervals in (\ref{a.18}). Since in the quantisation
of (\ref{a.1}) such a choice of $J$ would also have to be made, we simply
define it by to be the one chosen by the quantum commutator computation.
That is to say, we take the point of view spelled out in \cite{QSD} and 
consider the right hand side of the classical Poisson bracket between
the $\tilde{D}_+$, which 
involves structure functions, as a new operator which must be quantised 
to some extent by prescriptions and techniques  
independent of those for $\tilde{D}_+$.

This is as close as one 
can hope to see the correspondence between (\ref{a.1}) and (\ref{a.15}).
The unavoidable ambiguities in the quantisation caused by the discontinuity of 
the representation can be exploited in order to close the quantum algebra
including the structure functions. Also the right hand side of the classical
Poisson brackets has to be quantised in terms of finite diffeomorphisms 
because the generators do not exist. One should maybe call this a ``soft
anomaly'', i.e. the fact that the classical computation is only reproduced 
by the finite approximations to the infinitesimal classical counterparts.
However, it is not a troublesome anomaly in the
sense that the quantum algebra still closes with the correct factor ordering
so that the solutions to the constraints do not have to obey any
extra properties. It is just that 
that the structure functions are replaced by somewhat deformed quantum 
operators due to the unavaoidable
quantisation ambiguity in the choice of $(\gamma,v)\mapsto 
\varphi_{\gamma,v}$. A similar ``soft" anomaly has been observed 
in 2+1 gravity with a cosmological constant \cite{Perez}.\\
\\
\\
We now turn to the $[\tilde{D}_+,\tilde{D}_-]$ commutator
\ba \label{a.19}
&& [\tilde{D}_+[f],\tilde{D}_-[g]]\; |\gamma^+,\gamma^->
\nonumber\\
&=& 
\sum_{v\in V(\gamma^-)}\; g(v)\; \lambda(\gamma^+,\gamma^-,v)\;
D_+[f]\;[|\gamma^+,\gamma^-_v> -|\gamma^+,\gamma^->]
\nonumber\\
&& - 
\sum_{v\in V(\gamma^+)}\; f(v)\; \lambda(\gamma^+,\gamma^-,v)\;
D_-[g]\;[|\gamma^+_v,\gamma^-> -|\gamma^+,\gamma^->]
\nonumber\\
&=& 
\sum_{v\in V(\gamma^-)}\; g(v)\; \lambda(\gamma^+,\gamma^-,v)\;
\{\sum_{\tilde{v}\in V(\gamma^+)} \; f(\tilde{v})\; 
\lambda(\gamma^+,\gamma^-_v,\tilde{v})
[|\gamma^+_{\tilde{v}},\gamma^-_v>-|\gamma^+,\gamma^-_v>]
\nonumber\\
&&
-\sum_{\tilde{v}\in V(\gamma^+)} \; f(\tilde{v})\; 
\lambda(\gamma^+,\gamma^-,\tilde{v})
[|\gamma^+_{\tilde{v}},\gamma^->-|\gamma^+,\gamma^->]\}
\nonumber\\
&&
-\sum_{v\in V(\gamma^+)}\; f(v)\; \lambda(\gamma^+,\gamma^-,v)\;
\{\sum_{\tilde{v}\in V(\gamma^-)} \; g(\tilde{v})\; 
\lambda(\gamma^+_v,\gamma^-,\tilde{v})
[|\gamma^+_v,\gamma^-_{\tilde{v}}>-|\gamma^+_v,\gamma^->]
\nonumber\\
&&
-\sum_{\tilde{v}\in V(\gamma^-)} \; g(\tilde{v})\; 
\lambda(\gamma^+,\gamma^-,\tilde{v})
[|\gamma^+,\gamma^-_{\tilde{v}}>-|\gamma^+,\gamma^->]\}
\ea
Relabelling $v\leftrightarrow \tilde{v}$ in the second term gives
\ba \label{a.19a}
&& [\tilde{D}_+[f],\tilde{D}_-[g]]\; |\gamma^+,\gamma^->
\nonumber\\
&=& 
\sum_{v\in V(\gamma^-),\tilde{v}\in V(\gamma_+)}\; 
g(v)\; \lambda(\gamma^+,\gamma^-,v)\; f(\tilde{v})\;
\times\nonumber\\ && 
\{\lambda(\gamma^+,\gamma^-_v,\tilde{v})\;
[|\gamma^+_{\tilde{v}},\gamma^-_v>-|\gamma^+,\gamma^-_v>]
-\lambda(\gamma^+,\gamma^-,\tilde{v})
[|\gamma^+_{\tilde{v}},\gamma^->-|\gamma^+,\gamma^->]\}
\nonumber\\
&&
-\sum_{v\in V(\gamma^-),\tilde{v}\in V(\gamma^+)}\; 
f(\tilde{v})\; \lambda(\gamma^+,\gamma^-,\tilde{v})\; g(v)\; 
\times\nonumber\\ && 
\{\lambda(\gamma^+_{\tilde{v}},\gamma^-,v)
[|\gamma^+_{\tilde{v}},\gamma^-_v>-|\gamma^+_{\tilde{v}},\gamma^->]
-\lambda(\gamma^+,\gamma^-,v)
[|\gamma^+,\gamma^-_v>-|\gamma^+,\gamma^->]\}
\nonumber\\
&=& 
\sum_{v\in V(\gamma^-),\tilde{v}\in V(\gamma_+}\; g(v)\; f(\tilde{v})
\{
[\lambda(\gamma^+,\gamma^-,v)\lambda(\gamma^+,\gamma^-_v,\tilde{v})
-\lambda(\gamma^+,\gamma^-,\tilde{v})
\lambda(\gamma^+_{\tilde{v}},\gamma^-_v,v)]\;
|\gamma^+_{\tilde{v}},\gamma^-_v>
\nonumber\\
&&
+\lambda(\gamma^+,\gamma^-,v)[-\lambda(\gamma^+,\gamma^-_v,\tilde{v})
+\lambda(\gamma^+,\gamma^-,\tilde{v})]\;|\gamma^+,\gamma^-_v>
\nonumber\\
&&
+\lambda(\gamma^+,\gamma^-,\tilde{v})[
-\lambda(\gamma^+,\gamma^-,\tilde{v})
+\lambda(\gamma^+_{\tilde{v}},\gamma^-,v)]\;
|\gamma^+_{\tilde{v}},\gamma^->
\}
\ea
Recall that $\lambda(\gamma^+,\gamma^-,v)\not=0$ only if 
$v\in V(\gamma^+)\cap V(\gamma^-)$ so that sum over $v,\tilde{v}$ runs over
the same effective range. Furthermore, for $\tilde{v}\not=v$ 
we have 
\be \label{a.20}
\lambda(\gamma^+,\gamma^-_v,\tilde{v})=\lambda(\gamma^+,\gamma^-_v,\tilde{v}),
\;\; 
\lambda(\gamma^+_{\tilde{v}},\gamma^-,v)=\lambda(\gamma^+,\gamma^-,v)
\ee
Therefore the double sum in the second and third term collapses
to $\tilde{v}=v$. In the first term, the contribution for $\tilde{v}\not=v$
again vanishes due to (\ref{a.20}) while for $v=\tilde{v}$ we have 
$\lambda(\gamma^+,\gamma^-_v,\tilde{v})=
\lambda(\gamma^+_{\tilde{v}},\gamma^-,v)=0$
identically because $v$ is no vertex of $\gamma^\pm_v$. 
Therefore the contribution from the first term vanishes and we are left with 
\ba \label{a.21}
&& [\tilde{D}_+[f],\tilde{D}_-[g]]\; |\gamma^+,\gamma^->
\nonumber\\
&=& \sum_{v\in V(\gamma^+)\cap V(\gamma^-)}\; f(v)\; g(v)\; 
\lambda(\gamma^+,\gamma^-,v)^2\{
[|\gamma^+,\gamma^-_v>-|\gamma^+,\gamma^->]-
[|\gamma^+_v,\gamma^->-|\gamma^+,\gamma^->]\}
\nonumber\\
&=& \sum_{v\in V(\gamma^+)\cap V(\gamma^-)}\; f(v)\; g(v)\; 
\{[U(\varphi^+_v,1)-1]-[U(1,\varphi^-_v)-1]\}
Q(v)^2\;|\gamma^+,\gamma^->
\ea
Comparing with (\ref{a.2}) we see to some extent the correct structure:
The result is local in that only a single sum is involved, it is a linear
combination of an infinitesimal $D_+$ and $D_-$ constraint. However, 
in contrast to the classical computation no (discrete) derivatives 
seem to appear and the structure functions do not 
seem to match the $Q(v)^2$ operator. Yet, we can match the two expressions 
as follows:\\
Recall the end result 
\be \label{a.22}
\{\tilde{D}_+[f],\tilde{D}_-[g]\}=
-\frac{1}{2}\int\;dx\; f\; D_+\; \frac{X'_+}{\sqrt{\det(q)}^3}
[g \frac{X'_-}{\sqrt{\det(q)}}]'
+\frac{1}{2}\int\;dx\; g\; D_-\; \frac{X'_-}{\sqrt{\det(q)}^3}
[f \sqrt{X'_+}{\sqrt{\det(q)}}]'
\ee
To quantise this expression we use the identity 
\be \label{a.23}
\{V(I_x),P_\pm(I'_x)\}=-\frac{1}{2}\frac{X_\mp'(x)}{\sqrt{\det(q)}(x)}
\ee
where $I_x$ is any interval containing in $x$ and $I'_x=[y,x]$ where 
$y\not\in I_x$. Thus 
\ba \label{a.24}
\{\tilde{D}_+[f],\tilde{D}_-[g]\} &=&
-2\int\;dx\; f\; D_+\; \frac{\{V_{I_x},P_-(I'_x)\}}{\det(q)}
[g \{V(I_x),P_+(I'_x)\}]'
\nonumber\\  &&
+2\int\;dx\; g\; D_-\; \frac{\{V_{I_x},P_+(I'_x)\}}{\det(q)}
[f \{V(I_x),P_-(I'_x)\}]'
\ea
As for the quantisation of $\tilde{D}_\pm$ we introduce a partition $\tau$
and replace the derivative by a difference using the intervals $J$ and their 
duals $J'$ of the 
triangulation 
\ba \label{a.25}
&& \{\tilde{D}_+[f],\tilde{D}_-[g]\} =
2\lim_{\tau\to S^1}\; \sum_{J\in \tau}\;
\times \nonumber\\
&&\{-f(b_J)\; D_+[J,J']
\frac{\{V(J),P_-(J')\}}{V(J)^2}
[g(f_J) \{V(J+1),P_+(J'+1)\}-g(b_J)\{V(J),P_+(J')\}]
\nonumber\\ &&
+g(b_J)\; D_-[J,J']
\frac{\{V(J),P_+(J')\}}{V(J)^2}
[f(f_J) \{V(J+1),P_-(J'+1)\}-f(b_J)\{V(J),P_-(J')\}]
\}
\ea
where $J+1,J'+1$ is the (dual) interval next neighbour to $J,J'$ respectively.
Now the inverse volume 
factors $1/V(J)$ can be treated as in section (\ref{s3}) giving
rise to $Q(J)^2$ so that
\ba \label{a.26}
&& \{\tilde{D}_+[f],\tilde{D}_-[g]\} =
2\lim_{\tau\to S^1}\; \sum_{J\in \tau}\;
\times \\ &&
\{-f(b_J)\; D_+[J,J']\; (\{V(J),P_-(J')\}
[g(f_J) \{V(J+1),P_+(J'+1)\}-g(b_J)\{V(J),P_+(J')\}])
\nonumber\\ &&
+g(b_J)\; D_-[J,J']\;(\{V(J),P_+(J')\}\;
[f(f_J) \{V(J+1),P_-(J'+1)\}-f(b_J)\{V(J),P_-(J')\}])
\}Q(J)^2
\nonumber
\ea
Now notice that 
\be \label{a.27}
-\{V(J),P_+(J')\}\;\{V(J),P_-(J')\}=1/4
\ee
so the second terms within the round brackets containing
the difference can be replaced by $1/4$. Now 
upon quantisation (\ref{a.26}) reduces to a sum over vertices and the correct 
$Q(v)^2$ operators and the correct infinitesimal $D_\pm$ appear. The
operator corresponding to $\{V(J+1),P_\pm(J'+1)\}$ in the term corresponding to 
vertex $v$ will then be 
$T^\pm_{I'_v+1,k_\pm=-k_0}[V(I_v+1),T^\pm_{I'_v+1,k_\pm=k_0}]$ where 
$I_v+1$ and $I'_v+1$ are (dual) intervals infinitesimally translated 
from $I_v,I'_v$, in particular they do not contain any vertex of the graphs.
Consequently its contribution vanishes and the operator indeed reduces to 
(\ref{a.21}) up to a factor of $1/2$ which, however, could be absorbed into 
the choices of the actual intervals that enter the definition of 
$D_\pm[J,J']$ and which cannot be fixed by the arguments of section
\ref{s3}. Hence also the $\tilde{D}_+,\tilde{D}_+$ commutator closes in the 
expected way up to a soft anomaly.

\end{appendix}

\end{document}